\documentclass[twocolumn,tighten]{aastex62}
\hypersetup{linkcolor=red,citecolor=blue,filecolor=blue,urlcolor=magenta}
\usepackage{color}

\def\deg{{$^{\circ}$}}
\def\bdp{\object[BD+03 740]{BD$+$03\deg740}}
\def\bdm{\object[BD-13 3442]{BD$-$13\deg3442}}
\def\cdm{\object[CD-33 1173]{CD$-$33\deg1173}}
\def\gsix{\object[G 64-12]{G~64-12}}
\def\hdone{\object[HD 19445]{HD~19445}}
\def\hdeight{\object[HD 84937]{HD~84937}}
\def\hdnine{\object[HD 94028]{HD~94028}}
\def\kmsec{\mbox{km~s$^{\rm -1}$}}
\def\logg{\mbox{log~{\it g}}}
\def\msun{\mbox{$M_{\odot}$}}
\def\teff{\mbox{$T_{\rm eff}$}}
\def\vt{\mbox{$v_{\rm t}$}}

\def\loggf{\mbox{$\log gf$}}

\shorttitle{Iron Abundances in Metal-Poor Dwarf Stars}
\shortauthors{I.U.\ Roederer et al.}
\accepted{for publication in the Astrophysical Journal}

\begin{document}

\title{
Consistent Iron Abundances Derived from Neutral and Singly-Ionized
Iron Lines \\ in Ultraviolet and Optical Spectra 
of Six Warm Metal-Poor Stars\footnote{%
Based on observations made with the NASA/ESA 
\textit{Hubble Space Telescope}, 
obtained at the Space Telescope Science Institute (STScI), which is 
operated by the Association of Universities for 
Research in Astronomy, Inc.\ (AURA) under NASA contract NAS~5-26555.
These observations are associated with program GO-14232.
Some
data presented in this paper were obtained from the 
Barbara A.\ Mikulski Archive for Space Telescopes (MAST).~
These data are associated with Programs GO-7402, GO-9049, and GO-14161.
Other data have been obtained from the European Southern Observatory (ESO) 
Science Archive Facility.
These data are associated with Programs
66.D-0636(A),
67.D-0439(A),
67.D-0554(A),
68.B-0475(A),
68.D-0094(A),
072.B-0585(A), and
095.D-0504(A).
This research has also made use of the Keck Observatory Archive (KOA), 
which is operated by the W.M.\ Keck Observatory and 
the NASA Exoplanet Science Institute (NExScI), 
under contract with NASA.
This work has also made use of data collected from
McDonald Observatory of the University of Texas at Austin.
} }

\email{Email: iur@umich.edu}

\author{Ian U.\ Roederer}
\affiliation{Department of Astronomy, University of Michigan,
1085 S.\ University Ave., Ann Arbor, MI 48109, USA}
\affiliation{Joint Institute for Nuclear Astrophysics - Center for the
Evolution of the Elements (JINA-CEE), USA}

\author{Christopher Sneden}
\affiliation{Department of Astronomy and McDonald Observatory,
The University of Texas, Austin, TX 78712, USA}

\author{James E.\ Lawler}
\affiliation{Department of Physics, University of Wisconsin-Madison,
Madison, WI 53706, USA}

\author{Jennifer S.\ Sobeck}
\affiliation{Department of Astronomy, University of Washington,
Seattle, WA 98195, USA}

\author{John J.\ Cowan}
\affiliation{Homer L.\ Dodge Department of Physics and Astronomy, 
University of Oklahoma, Norman, OK 73019, USA}

\author{Ann Merchant Boesgaard}
\affiliation{Institute for Astronomy, University of Hawai'i at Manoa, 
2680 Woodlawn Drive, Honolulu, HI 96822, USA}

\begin{abstract}

Neutral Fe lines in metal-poor stars yield
conflicting abundances depending on whether and how deviations from
local thermodynamic equilibrium (LTE) are considered.
We have collected new high resolution and high signal-to-noise
ultraviolet (UV) spectra of
three warm dwarf stars with [Fe/H]~$\approx -$2.9
with the Space Telescope Imaging Spectrograph on the 
\textit{Hubble Space Telescope}.
We locate archival UV spectra 
for three other warm dwarfs with [Fe/H]~$\approx -$3.3,
$-$2.2, and $-$1.6,
supplemented with
optical spectra for all six stars.
We calculate stellar parameters using methods that are largely independent
of the spectra, adopting broadband photometry, 
color-temperature relations, \textit{Gaia} parallaxes,
and assumed masses.
We use the LTE line analysis code MOOG to derive Fe abundances from
hundreds of Fe~\textsc{i} and Fe~\textsc{ii} lines
with wavelengths from 2290 to 6430~\AA.~
The [Fe/H] ratios derived separately
from Fe~\textsc{i} and Fe~\textsc{ii}
lines agree in all six stars, with
[Fe~\textsc{ii}/H]~$-$~[Fe~\textsc{i}/H] 
ranging from
$+$0.00~$\pm$~0.07 to
$-$0.12~$\pm$~0.09~dex,
when strong lines and
Fe~\textsc{i} lines with lower excitation potential
$<$~1.2~eV are excluded.
This constrains the extent of any
deviations from LTE that may occur
within this parameter range.
While our result confirms
non-LTE calculations for some warm, metal-poor dwarfs,
it may not be generalizable to
more metal-poor dwarfs,
where deviations from LTE are predicted to be larger.
We also investigate trends of
systematically lower abundances derived from
Fe~\textsc{i} lines in the Balmer continuum region
($\approx$~3100--3700~\AA),
and we conclude that no proposed explanation for this effect
can fully account for the observations presently available.

\end{abstract}


\section{Introduction}
\label{intro}

The iron-group elements (21~$\leq Z \leq$~30) observed
in metal-poor stars were produced in
massive-star supernovae early in the
history of the Galaxy.
Supernova models predict yields for individual elements
as a function of initial mass, metallicity, 
rotation rate, and explosion physics.
The predicted elemental abundance patterns can be incorporated into 
models of the chemical evolution of the Galaxy
or individual dwarf galaxies, where the yields,
initial mass function, and supernova models
are tuned and refined to 
reproduce the observed abundance patterns
(e.g., \citealt{nomoto13}).

Yet concerns remain that the observed abundance patterns
of the iron-group elements may not always be reliable.
The singly-ionized species of iron-group elements dominate 
by number ($\gtrsim$~95\% for Sc through Cu; 
e.g., \citealt{sneden16}) 
in the line-forming layers of the
atmospheres of late-type (FGK) stars.
However, most of the lines detected in optical spectra
accessible from the ground (3020~$< \lambda \lesssim$~10,000~\AA)
arise from the neutral, minority species.
This situation is further aggravated 
in the most metal-poor stars ([Fe/H]~$< -$2.5),
because all metal lines get progressively weaker,
and there are few if any lines of
singly-ionized iron-group elements available
in the optical.
Lines from the neutral species may not always yield 
reliable abundances, because the
assumption of local thermodynamic equilibrium (LTE) 
may be invalid
when computing excitation and ionization equilibria
in late-type, metal-poor stars
(e.g., \citealt{thevenin99,korn03,collet05,
mashonkina11,bergemann12,lind12,ezzeddine17}).

This issue can be avoided by deriving abundances 
from weak lines connected to the ground and low-lying
levels of the majority, singly-ionized species.
These levels contain the bulk of the populations,
so they should
not deviate appreciably from their LTE
Boltzmann and Saha equilibria values.
Lines of these species are plentiful in
the near-ultraviolet (UV) portion of the spectrum
(2000~$\lesssim \lambda \lesssim$~3100~\AA).~
This region of the spectrum is reasonably uncrowded
in warm, metal-poor dwarf stars,
so individual lines
and the local continuum level can be discerned with confidence.
Stars like these are ideal laboratories to test
the reliability and applicability of LTE calculations
to the analysis of iron-group abundances in metal-poor stars.

We adopted this approach in our pilot study of
the warm, metal-poor dwarf \hdeight.
\citet{sneden16} conducted an analysis of
446 Fe~\textsc{i} lines and 105 Fe~\textsc{ii} lines in \hdeight,
finding that the [Fe/H] ratios computed independently
from the two species agreed to better than 0.01~dex,
with a total systematic uncertainty less than 0.1~dex.
\citeauthor{sneden16}\ adopted model parameters for \hdeight\
that were calculated independent of the spectra (see \citealt{lawler13}).
This is an important point, because many abundance analyses
of metal-poor stars lacking accurate parallax measurements
rely on the ionization equilibrium of Fe~\textsc{i} and \textsc{ii}
to determine the surface gravity.
By construction, this approach forces both Fe~\textsc{i} and 
Fe~\textsc{ii} lines to yield the same [Fe/H],
so it cannot be used to estimate possible non-LTE effects
in Fe~\textsc{i}.
Furthermore, it may bias the derived
surface gravity if Fe~\textsc{i} is not in LTE
(e.g., \citealt{lind12,sitnova15}).
\citeauthor{sneden16}\ did not enforce ionization equilibrium,
so their finding that both Fe~\textsc{i} and \textsc{ii} lines
yield the same [Fe/H] ratio is an independent confirmation 
that the deviations from LTE cannot be too severe.
\citeauthor{sneden16}\ also analyzed 152
Fe~\textsc{i} and 16 Fe~\textsc{ii} lines in the Sun,
where Fe~\textsc{i} and Fe~\textsc{ii} lines each
yielded [Fe/H] ratios that agreed at the 0.01~dex level.

As a natural extension of this study,
we analyze six additional bright, metal-poor stars near
the main-sequence turnoff point.
Three of these stars with [Fe/H]~$\approx -$2.9,
\bdp, \bdm, and \cdm, form the core of the project
for which new UV observations were obtained.
Three other stars
with high-quality archival UV spectra, 
\gsix, \hdone, and \hdnine,
are included to provide benchmarks 
across a wider metallicity range
($-$3.3~$\leq$~[Fe/H]~$\leq -$1.6)
for abundances used in chemical evolution models.
We also collect new and archival high-resolution
optical spectra of all six stars.
These observations are described in detail in 
Section~\ref{obs}.
We discuss our line list and transition probabilities
in Section~\ref{linelist}, 
and
we validate our data reduction by
comparing equivalent widths (EWs)
with previous work in Section~\ref{ew}.
We calculate stellar parameters using methods
that are largely independent of the spectra
in Section~\ref{params}.
This enables us to independently assess
the [Fe/H] values derived separately from Fe~\textsc{i}
and Fe~\textsc{ii} lines.
We compare our derived Fe abundances with
our previous work on \hdeight\
in Section~\ref{hd84937}.
We discuss our results in Section~\ref{discussion}
and summarize our conclusions in Section~\ref{conclusions}.
Our complete analysis of other iron-group element abundances,
and their implications for supernova yields
and Galactic chemical evolution,
will be presented in future work.

We adopt the standard definitions of elemental abundances and ratios.
For element X, the logarithmic absolute abundance is defined
as the number of atoms of X per 10$^{12}$ hydrogen atoms,
$\log\epsilon$(X)~$\equiv \log_{10}(N_{\rm X}/N_{\rm H}) +$12.0.
For elements X and Y, the logarithmic abundance ratio relative to the
Solar ratio is defined as
[X/Y]~$\equiv \log_{10} (N_{\rm X}/N_{\rm Y}) -
\log_{10} (N_{\rm X}/N_{\rm Y})_{\odot}$.
We adopt the Solar Fe abundance of \citet{asplund09},
$\log\epsilon$(Fe)~$=$~7.50.
Abundances or ratios denoted with the ionization state
are defined to be 
the total elemental abundance as derived from transitions of
that particular ionization state 
after Saha ionization corrections have been applied.

\section{Observations}
\label{obs}

Our analysis is based on several new sets of
UV spectra, supplemented with new or archival
UV and optical spectra.
These spectra cover most of the 
near-UV and optical spectral ranges
of each of the six stars.
Table~\ref{obstab} lists some key features 
of the spectra, including
the instrument, program identification (ID) number,
principle investigator (PI), 
wavelength range covered ($\lambda$), 
spectral resolving power ($R \equiv \lambda/\Delta\lambda$),
and signal-to-noise (S/N) ratio at a 
representative wavelength.

\begin{deluxetable*}{lcccccc}
\tablecaption{Characteristics of Near-UV and Optical Spectra
\label{obstab}}
\tablewidth{0pt}
\tabletypesize{\scriptsize}
\tablehead{
\colhead{Star} &
\colhead{Instrument} &
\colhead{Program ID} &
\colhead{PI} &
\colhead{$\lambda$ (\AA)} &
\colhead{$R$} &
\colhead{S/N pix$^{-1}$@$\lambda$}
}
\startdata
\bdp         & STIS  & GO-14232\tablenotemark{a} & Roederer                   &  2278--3068 & 30,000 &  70@2820~\AA \\
($V =$~9.81) & UVES  & 68.D-0094(A)              & Primas                     &  3050--3869 & 41,000 & 230@3500~\AA \\
             & HIRES & H41aH                     & Boesgaard\tablenotemark{i} &  3568--3950 & 47,000 & 280@3650~\AA \\
             & Tull  & \nodata                   & Sneden                     & 3680--10120 & 33,000 & 180@5100~\AA \\
             & HIRES & U10H                      & Bolte\tablenotemark{j}     &  3970--4917 & 47,000 & 200@4500~\AA \\
             & UVES  & 68.D-0094(A)              & Primas                     &  4780--5755 & 51,000 & 480@5100~\AA \\
             & UVES  & 68.D-0094(A)              & Primas                     &  5834--6804 & 51,000 & 320@6000~\AA \\
\hline
\bdm         & STIS  & GO-14232\tablenotemark{b} & Roederer                   &  2278--3069 & 30,000 &  60@2820~\AA \\
($V =$~10.27)& UVES  & 67.D-0439(A)              & Primas\tablenotemark{k}    &  3060--3867 & 49,000 & 125@3500~\AA \\
             & UVES  & 095.D-0504(A)             & Melendez\tablenotemark{l}  &  3300--4515 & 49,000 & 110@4000~\AA \\
             & UVES  & 67.D-0439(A)              & Primas\tablenotemark{k}    &  4785--5755 & 57,000 & 230@5100~\AA \\
             & UVES  & 67.D-0439(A)              & Primas\tablenotemark{k}    &  5833--6805 & 57,000 & 210@6000~\AA \\
\hline
\cdm         & STIS  & GO-14232\tablenotemark{c} & Roederer                   &  2277--3069 & 30,000 &  55@2820~\AA \\
($V =$~10.90)& UVES  & 68.B-0475(A)              & Primas                     &  3050--3867 & 37,000 & 160@3500~\AA \\
             & UVES  & 095.D-0504(A)             & Melendez                   &  3305--4520 & 49,000 &  90@4000~\AA \\
             & UVES  & 68.B-0475(A)              & Primas                     &  4887--5755 & 46,000 & 200@5100~\AA \\
             & UVES  & 68.B-0475(A)              & Primas                     &  5832--6806 & 46,000 & 190@6000~\AA \\
\hline
\gsix        & STIS  & GO-9049\tablenotemark{d}  & Deliyannis                 &  2001--2812 & 30,000 &  25@2820~\AA \\
($V =$~11.45)& HIRES & H11aH                     & Boesgaard\tablenotemark{i} &  3070--3903 & 49,000 & 250@3500~\AA \\
             & UVES  & 67.D-0554(A)              & Christlieb                 &  3297--4490 & 41,000 & 150@4000~\AA \\
             & UVES  & 67.D-0554(A)              & Christlieb                 &  4810--5727 & 42,000 & 250@5100~\AA \\ 
             & UVES  & 67.D-0554(A)              & Christlieb                 &  5824--6795 & 42,000 & 240@6000~\AA \\ 
\hline
\hdone       & STIS  & GO-7402\tablenotemark{e}  & Peterson\tablenotemark{m}  &  2313--3067 & 30,000 &  65@2820~\AA \\
($V =$~8.06) & UVES  & 68.D-0094(A)              & Primas\tablenotemark{k}    &  3055--3874 & 41,000 & 200@3500~\AA \\
             & UVES  & 66.D-0636(A)              & Piotto\tablenotemark{n}    &  3760--4980 & 41,000 & 220@4500~\AA \\
             & UVES  & 68.D-0094(A)              & Primas\tablenotemark{k}    &  4790--5760 & 51,000 & 300@5100~\AA \\
             & UVES  & 68.D-0094(A)              & Primas\tablenotemark{k}    &  5841--6810 & 51,000 & 200@6000~\AA \\
\hline
\hdeight     & STIS  & GO-7402\tablenotemark{f}  & Peterson\tablenotemark{m}  &  2279--3117 & 30,000 &  55@2820~\AA \\
($V =$~8.32) & STIS  & GO-14161\tablenotemark{g} & Peterson\tablenotemark{o}  &  2128--3143 &114,000 &  50@2820~\AA \\
\hline
\hdnine      & STIS  & GO-7402\tablenotemark{f}  & Peterson\tablenotemark{m}  &  2280--3117 & 30,000 &  40@2820~\AA \\
             & STIS  & GO-14161\tablenotemark{h} & Peterson\tablenotemark{o}  &  2128--3143 &114,000 &  40@2820~\AA \\
($V =$~8.22) & UVES  & 072.B-0585(A)             & Silva                      &  3050--3860 & 37,000 &  70@3500~\AA \\
             & Tull  & \nodata                   & Roederer\tablenotemark{p}  &  3650--8000 & 33,000 & 140@5100~\AA \\
\enddata
\tablenotetext{a}{Datasets OCTS01010, OCTS02010-30}
\tablenotetext{b}{Datasets OCTS03010-30, OCTS04010-20, OCTS05010-20}
\tablenotetext{c}{Datasets OCTS06010-30, OCTS07010-30, OCTS08010-30}
\tablenotetext{d}{Datasets O6ED01010-20, O6ED02010-20, O6ED03010-20, O6ED04010-20}
\tablenotetext{e}{Datasets 056D01010, 056D03010}
\tablenotetext{f}{Dataset 056D06010}
\tablenotetext{g}{Datasets OCTKA6010-D020}
\tablenotetext{h}{Datasets OCTKB0010-6030}
\tablenotetext{i}{See \citet{rich09}}
\tablenotetext{j}{See \citet{lai08}}
\tablenotetext{k}{See \citet{hansen12}}
\tablenotetext{l}{See \citet{reggiani17}}
\tablenotetext{m}{See \citet{peterson01}}
\tablenotetext{n}{See \citet{recioblanco02}}
\tablenotetext{o}{See \citet{peterson17}}
\tablenotetext{p}{See \citet{roederer14}}
\end{deluxetable*}

\begin{figure*}
\begin{center}
\includegraphics[angle=0,width=6.4in]{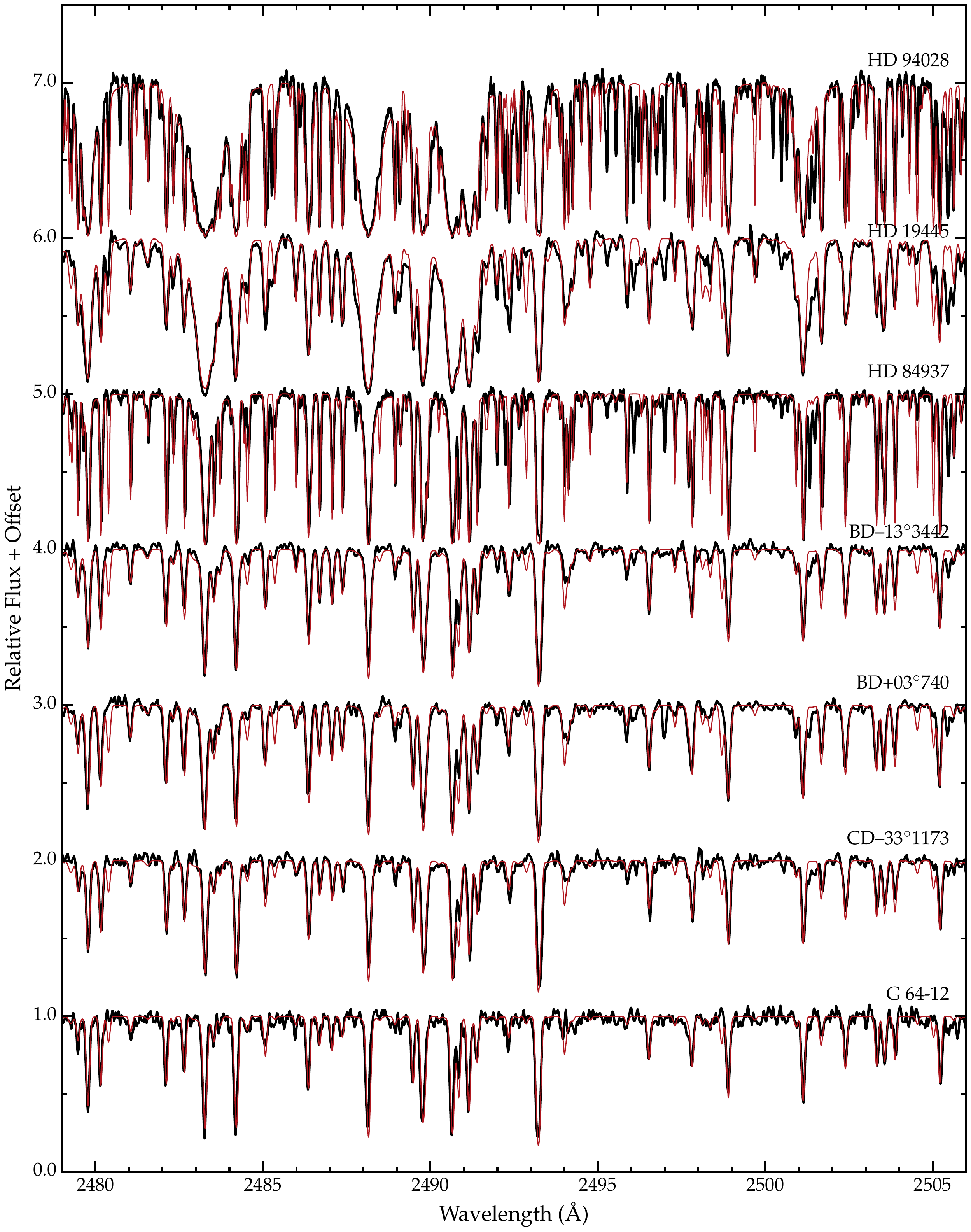}
\end{center}
\caption{
\label{specplot}
Portions of the STIS UV spectra of the six stars in our sample,
plus \mbox{HD~84937}.
The spectra are normalized to unity and offset vertically
to enhance visibility.
The bold black lines mark the observed spectra, and
the thin red lines mark the synthetic spectra.
The overwhelming majority of lines in this region are
due to neutral and singly-ionized species of
Fe group elements (Sc through Ni).
We adopt abundance ratios ([X/Fe])
for these iron-group elements, X,
from previous work \citep{aoki06,lai08,roederer14,sneden16,reggiani17}
and averaged together these ratios in cases with
no previous values.
Note that the spectra of 
\mbox{HD~84937} and \mbox{HD~94028} were taken with
STIS/E230H ($R \sim$~114,000), and the other spectra
were taken with STIS/E230M ($R \sim$~30,000);
see Section~\ref{highresolution} for details.
}
\end{figure*}

The new UV spectra were obtained using the
Space Telescope Imaging Spectrograph (STIS;
\citealt{kimble98,woodgate98}) on the
\textit{Hubble Space Telescope} (\textit{HST}).~
These observations were made using the E230M
echelle grating centered at $\lambda$2707, 
the 0\farcs06~$\times$~0\farcs2 slit,
the near-UV Multianode Microchannel Array (MAMA) detector.
Three stars were observed with STIS as part of Program GO-14232:\
\bdp, \bdm, and \cdm.
\bdp\ was observed over the course of
four orbits, or 9,556~s of integration time,
on 2016 Feb 16 and 2016 Mar 27.
\bdm\ was observed over the course of 
seven orbits, or 17,264~s of integration time,
on 2016 Dec 22 and 2016 Dec 24.
\cdm\ was observed over the course of 
nine orbits, or 23,355~s of integration time,
on 2016 Dec 31, 2017 Jan 01, 
and 2017 Jan 06-08.
The total counts in these observations generally
increase by factors of 2--3 from the shortest wavelengths
to $\sim$~2800~\AA,
where the reference S/N is listed in Table~\ref{obstab}.
Figure~\ref{specplot} illustrates a portion of the
UV spectrum of these stars.

Additional spectra 
are obtained from a variety of online archives,
including the Mikulski Archive for Space Telescopes (MAST),
the European Southern Observatory (ESO) Science Archive Facility,
and the Keck Observatory Archives.
These include data collected using the 
Ultraviolet and Visual Echelle Spectrograph (UVES; \citealt{dekker00})
at the Very Large Telescope (VLT UT2) and
the High Resolution Echelle Spectrometer (HIRES; \citealt{vogt94})
at the Keck~I Telescope.
The footnotes of Table~\ref{obstab} reference
publications by the original investigators
that describe these data in detail.

We also make use of spectra obtained using the 
Robert G.\ Tull Coud\'{e} Spectrograph \citep{tull95} at the 
Harlan J.\ Smith Telescope at McDonald Observatory.
The Tull spectrum of \hdnine\ has been described previously
by \citet{roederer14}.
The Tull spectrum of \bdp\ was obtained 
on 2016 Oct 21, with an integration time of 2100~s,
using an identical setup and following similar 
calibration and reduction procedures.

\section{Line List and Transition Probabilities}
\label{linelist}

We prioritize \loggf\ values derived
from modern laboratory studies
that use
radiative lifetime measurements 
from laser-induced fluorescence
and emission branching fraction data 
from high resolution spectrometers.
This approach routinely achieves high
precision absolute \loggf\ values whose uncertainties
are $\approx$~5\% or better.
We adopt an initial Fe~\textsc{i} line list from
\citet{sneden16}, who made use of laboratory 
\loggf\ values published by
\citet{obrian91}, via the 
National Institute of Standards and Technology (NIST)
Atomic Spectral Database (ASD) \citep{kramida17},
and recent work by 
\citet{denhartog14} and \citet{ruffoni14}.
The median uncertainty in the \loggf\ values for the
lines measured in at least one star
in our sample is $\approx$~7\% (NIST grade B+),
or $\approx$~0.03~dex.

Our initial Fe~\textsc{ii} line list 
also comes from \citet{sneden16},
who adopted the \loggf\ values from NIST.~
Modern laboratory work on Fe~\textsc{ii} lines is not as
extensive as for Fe~\textsc{i}, and 
the median uncertainty of the lines in our list
is $\approx$~25\% (NIST grade C),
or $\approx$~0.12~dex.
The NIST values are adopted from many studies, including
\citet{bergeson96}, \citet{sikstrom99},
\citet{pickering01,pickering02}, and \citet{schnabel04}.
Comparisons by \citeauthor{bergeson96}\ between their results
and values in the
NIST ASD at that time indicated that 
no serious systematic differences were present.
The list of Fe~\textsc{i} and \textsc{ii} lines
measured in at least one star in our sample
is given in Table~\ref{ewtab}.

\startlongtable
\begin{deluxetable*}{cccccccccccccccccc}
\tablecaption{Fe Line List, Equivalent Widths, and Abundances
\label{ewtab}}
\tablewidth{0pt}
\tabletypesize{\scriptsize}
\tablehead{
\colhead{Species} &
\colhead{Wavelength} &
\colhead{E.P.} &
\colhead{\loggf} &
\colhead{EW} &
\colhead{EW} &
\colhead{EW} &
\colhead{EW} &
\colhead{EW} &
\colhead{EW} &
\colhead{EW} &
\colhead{$\log\epsilon$} &
\colhead{$\log\epsilon$} &
\colhead{$\log\epsilon$} &
\colhead{$\log\epsilon$} &
\colhead{$\log\epsilon$} &
\colhead{$\log\epsilon$} &
\colhead{$\log\epsilon$} \\
\colhead{} &
\colhead{(\AA)} &
\colhead{(eV)} &
\colhead{} &
\colhead{(m\AA)} &
\colhead{(m\AA)} &
\colhead{(m\AA)} &
\colhead{(m\AA)} &
\colhead{(m\AA)} &
\colhead{(m\AA)} &
\colhead{(m\AA)} &
\colhead{} &
\colhead{} &
\colhead{} &
\colhead{} &
\colhead{} &
\colhead{} &
\colhead{} \\
\colhead{} &
\colhead{} &
\colhead{} &
\colhead{} &
\colhead{[1]} &
\colhead{[2]} &
\colhead{[3]} &
\colhead{[4]} &
\colhead{[5]} &
\colhead{[6]} &
\colhead{[7]} &
\colhead{[1]} &
\colhead{[2]} &
\colhead{[3]} &
\colhead{[4]} &
\colhead{[5]} &
\colhead{[6]} &
\colhead{[7]} 
}
\startdata
 Fe~\textsc{i}  & 2297.79 & 0.05 & $-$1.10 &\nodata&\nodata&  53.3 &\nodata&\nodata&\nodata&\nodata&\nodata&\nodata& 4.817 &\nodata&\nodata&\nodata&\nodata\\ 
 Fe~\textsc{i}  & 2369.46 & 0.11 & $-$2.19 &  20.6 &  22.7 &\nodata&\nodata&\nodata&\nodata&\nodata& 4.616 & 4.741 &\nodata&\nodata&\nodata&\nodata&\nodata\\ 
 Fe~\textsc{i}  & 2371.43 & 0.09 & $-$1.95 &  34.8 &  42.5 &  28.2 &  23.7 &  64.0 &\nodata&  75.5 & 4.749 & 5.055 & 4.843 & 4.591 & 5.589 &\nodata& 6.069 \\ 
\enddata
\tablecomments{[1] \bdp; [2] \bdm; [3] \cdm; [4] \gsix; [5] \hdone; [6] \hdeight; [7] \hdnine}
\tablecomments{The complete version of Table~\ref{ewtab} is available
in the online edition of the journal.  
An abbreviated version is shown here to illustrate its form and content.}
\end{deluxetable*}

Previous studies (e.g., \citealt{cayrel04,cohen08,cohen13,lai08})
have found that 
Fe~\textsc{i} lines 
with low excitation potentials (E.P.)\
may yield LTE abundances higher than average
in metal-poor dwarfs and giants.
\citet{bergemann12} confirmed this phenomenon
in their models
and advocated using Fe~\textsc{ii} 
and high-excitation Fe~\textsc{i} lines 
in determining metallicities.
We have many low-E.P.\ Fe~\textsc{i} lines in our list, and
we measure EWs for and derive abundances from these lines.
We report these values in Table~\ref{ewtab}, but
we exclude Fe~\textsc{i} lines with E.P.~$<$~1.2~eV 
from our adopted abundance calculations,
following recommendations from the studies referenced previously.
As we discuss in Section~\ref{agreement}, 
the Fe~\textsc{i} lines with E.P.~$<$~1.2~eV 
do yield higher abundances than the lines with
E.P.~$>$~1.2~eV in our study,
although the magnitude of the difference in the
mean is small ($\leq$~0.03~dex) in six of the seven stars.

\section{Equivalent Widths}
\label{ew}

We measure EWs using a semi-automatic 
routine that fits Voigt or Gaussian line profiles to 
continuum-normalized spectra at 
wavelengths of interest
(see \citealt{roederer14}).
Upon visual inspection of each line in each star, 
any line determined to be
undetected, blended, suffer from uncertain
continuum placement, or otherwise compromised
is discarded from consideration.
The EWs are listed in Table~\ref{ewtab}.

\subsection{Comparisons with Other Optical Spectra}

We deliberately obtain several sets of spectra of
\bdp\ that overlap in wavelength
to check the reproducibility of the EW measurements.
For a set of 49 lines in common between the 
UVES and Tull spectra of \bdp, we find a difference 
(UVES$-$Tull) of 
$-$0.5~$\pm$~0.3~m\AA\ ($\sigma =$~1.8~m\AA).~
For a set of 28 lines in common between the
HIRES and Tull spectra of \bdp, we find a difference
(HIRES$-$Tull) of 
$+$0.3~$\pm$~0.6~m\AA\ ($\sigma =$~3.3~m\AA).~
These differences are small and not significant.
We regard the spectra as interchangeable, although
we prioritize the one with higher S/N and 
resolving power in regions of overlap.

\subsection{Comparisons with Previous Work}

We also use \bdp\ to compare our EW measurements to
those made by other investigators.
Studies with at least 5 published EWs in common
with us are listed in Table~\ref{ewcomptab}.
The mean differences 
and standard deviations are small, indicating 
that our EWs are in good agreement with 
previous studies of this star.
These internal and external comparisons
support our assertion that our method of 
measuring EWs is reliable within the
range of stellar parameters and quality of the
spectra used in the present study.

\begin{deluxetable}{cccc}
\tablecaption{Comparison of Equivalent Width Measurements of
BD$+$03$^{\circ}$740 with Previous Studies
\label{ewcomptab}}
\tablewidth{0pt}
\tabletypesize{\scriptsize}
\tablehead{
\colhead{Previous study} &
\colhead{Mean} &
\colhead{Standard} &
\colhead{Number} \\
\colhead{} &
\colhead{difference\tablenotemark{a}} &
\colhead{deviation} &
\colhead{of lines} \\
\colhead{} &
\colhead{(m\AA)} &
\colhead{(m\AA)} &
\colhead{} }
\startdata
\citet{fulbright00} &    0.0 $\pm$ 0.7 & 2.1 &  8 \\
\citet{carretta02}  & $+$1.0 $\pm$ 0.3 & 1.2 & 24 \\
\citet{ivans03}     & $+$2.1 $\pm$ 0.7 & 3.8 & 27 \\
\citet{lai08}       & $-$0.7 $\pm$ 0.3 & 2.1 & 64 \\
\citet{hosford09}   & $+$2.0 $\pm$ 0.6 & 4.3 & 51 \\
\citet{rich09}      & $+$1.2 $\pm$ 0.6 & 2.6 & 19 \\
\enddata
\tablenotetext{a}{In the sense of 
EW$_{\rm previous}$ $-$ EW$_{\rm this~study}$}
\end{deluxetable}

\subsection{Comparisons with Higher Resolution UV Spectra}
\label{highresolution}

Higher-resolution archival STIS spectra covering the same
UV wavelength region are available 
for \hdeight\ and \hdnine,
as pointed out by the referee.
These spectra were taken using the E230H grating, 
which provides a resolving power $R \sim$~114,000.
We download, co-add, and continuum normalize these spectra
using the same approach used for the E230M $R \sim$~30,000 spectra.
Table~\ref{obstab} lists the characteristics of these spectra.

To determine what
impact the spectral resolution may have on our EW measurements,
we compare the EWs of UV lines measured in the
E230M spectra of \hdeight\ and \hdnine\ 
to those measured in the E230H spectra.
These values are listed in Table~\ref{ewhighrestab}.
For \hdeight, the mean difference in the EW measurements
(E230M minus E230H) is 
$+$0.5~$\pm$~0.5~m\AA\ ($\sigma =$~4.7~m\AA, 93~lines).
For \hdnine, the mean difference is
$-$0.9~$\pm$~2.3~m\AA\ ($\sigma =$~6.6~m\AA, 8 lines).
Neither difference is significant.

\startlongtable
\begin{deluxetable*}{ccccccccccc}
\tablecaption{Comparison of Fe Equivalent Widths and Abundances from E230M and E230H Spectra
\label{ewhighrestab}}
\tablewidth{0pt}
\tabletypesize{\scriptsize}
\tablehead{
\colhead{Star} &
\colhead{Species} &
\colhead{$\lambda$} &
\colhead{E.P.} &
\colhead{\loggf} &
\colhead{EW} &
\colhead{EW} &
\colhead{$\log\epsilon$} &
\colhead{$\log\epsilon$} &
\colhead{$\log\epsilon$} &
\colhead{$\log\epsilon$} \\
\colhead{} &
\colhead{} &
\colhead{(\AA)} &
\colhead{(eV)} &
\colhead{} &
\colhead{(m\AA)} &
\colhead{(m\AA)} &
\colhead{} &
\colhead{} &
\colhead{} &
\colhead{} \\
\colhead{} &
\colhead{} &
\colhead{} &
\colhead{} &
\colhead{} &
\colhead{[1]} &
\colhead{[2]} &
\colhead{[3]} &
\colhead{[4]} &
\colhead{[5]} &
\colhead{[6]} 
}
\startdata
HD84937 & FeI  & 2297.79  & 0.05   & $-$1.10  &  66.1 &  67.0  &  5.225 &  5.256 & 5.27 & 5.32  \\
HD84937 & FeI  & 2299.22  & 0.09   & $-$1.55  &  54.4 &  60.3  &  5.227 &  5.472 & 5.27 & 5.26  \\     
HD84937 & FeI  & 2369.46  & 0.11   & $-$2.19  &  33.1 &  38.1  &  5.073 &  5.222 & 5.12 & 5.24  \\     
\enddata
\tablecomments{%
[1] EW measured in E230M spectrum;
[2] EW measured in E230H spectrum;
[3] $\log\epsilon$ abundance derived from EW measured in E230M spectrum;
[4] $\log\epsilon$ abundance derived from EW measured in E230H spectrum;
[5] $\log\epsilon$ abundance derived from synthesis of E230M spectrum 
    (\citealt{sneden16} in the case of \mbox{HD~84937}, 
     this study in the case of \mbox{HD~94028});
[6] $\log\epsilon$ abundance derived from synthesis of E230H spectrum
}
\tablecomments{The complete version of Table~\ref{ewtab} is available
in the online edition of the journal.  
An abbreviated version is shown here to illustrate its form and content.}
\end{deluxetable*}

\section{Stellar Parameters}
\label{params}

We calculate the effective temperature (\teff)
and log of the surface gravity (\logg)
of each star using methods that are
largely independent of our spectra.
We use metallicity-dependent 
calibrations to calculate \teff\
from broadband colors,
but these calibrations have only a weak dependence on
metallicity for metal-poor stars.
The \logg\ value is calculated from two parameters
with mild metallicity dependences,
bolometric correction and \teff.
The microturbulent velocity (\vt)
and metallicity of the model atmosphere ([M/H]) 
are determined iteratively along with the
Fe abundance, so these parameters 
depend on the spectra.
The effects of [M/H] on \vt\ are expected to be small 
($<$~0.1~\kmsec), however, over the parameter
range of interest \citep{lind12}.

\subsection{Effective Temperatures}
\label{temperature}

We compile optical and near infrared broadband photometry
using catalogs from
\citet{ducati02} and \citet{munari14} 
for Johnson $B$ and $V$;  
\citet{cutri03} for 2MASS $J$, $H$, and $K$;
and
\citet{paunzen15} for Str\"{o}mgren-Crawford $b$ and $y$.
We construct six colors 
($B-V$, $V-J$, $V-H$, $V-K$, $J-K$, and $b-y$)
from these data.

We adopt reddening estimates from \citet{casagrande11},
when available, 
otherwise we modify the reddening predicted by the
\citet{schlafly11} dust maps.
These stars are nearby and in the foreground of some of
the reddening layer, so we reduce the \citeauthor{schlafly11}\ reddening
estimates by multiplying $E(B-V)$ by
$1 - \exp{(-|d \sin b|/h)}$
\citep{bonifacio00}.
Here, $d$ is the distance to the star (Section~\ref{gravity}),
and $b$ is its Galactic latitude.
Following \citeauthor{bonifacio00},
we assume a scale height, $h$, of 125~pc for
the reddening layer.
The $E(B-V)$ values from the \citeauthor{casagrande11}\ catalog
and the reduced \citeauthor{schlafly11} values are always small,
$<$~0.025.

We independently estimate the 
reddening using interstellar Na~\textsc{i} absorption visible
in the spectra.
The ratios of the EWs of the interstellar doublet lines 
at 5889.95 and 5895.92~\AA\
are $\approx$~2, which matches
the ratio of their $f$-values.
This indicates an optically thin column.
Following \citet{alvesbrito10} and \citet{roederer12}, 
we transform the EW into the Na~\textsc{i} column density 
as given by \citet{spitzer68}, 
the Na~\textsc{i} column density into the
H column density as given by \citet{ferlet85}, and the
H column density into $E(B-V)$ as given by \citet{bohlin78}.
The typical statistical uncertainties using this method are
$\approx$~0.002~mag, ignoring uncertainties in the calibrations themselves.
$E(B-V) \leq$~0.018 in all cases, in agreement with 
other estimates, as reported in Table~\ref{redtab}.
We deredden according to the extinction coefficients of \citet{mccall04}.

\begin{deluxetable}{lcccc}
\tablecaption{Reddening Estimates
\label{redtab}}
\tablewidth{0pt}
\tabletypesize{\scriptsize}
\tablehead{
\colhead{Star} &
\colhead{$E(B-V)$} &
\colhead{$E(B-V)$} &
\colhead{$E(B-V)$} &
\colhead{$E(B-V)$} \\
\colhead{} &
\colhead{[1]} &
\colhead{[2]} &
\colhead{[3]} &
\colhead{[4]} 
}
\startdata
\bdp\     & 0.00  & 0.020 & 0.022 & 0.018   \\
\bdm\     &\nodata& 0.025 & 0.011 & 0.012   \\
\cdm\     &\nodata& 0.010 & 0.005 & 0.003   \\
\gsix\    &\nodata& 0.021 & 0.003 & 0.008   \\
\hdone\   & 0.00  & 0.019 & 0.000 &$<$0.002 \\
\hdeight\ & 0.00  & 0.012 & 0.005 & 0.007   \\
\hdnine\  & 0.00  & 0.007 & 0.000 &$<$0.001 \\
\enddata
\tablecomments{%
[1] \citet{casagrande11} color calibrations; 
[2] \citet{schlafly11} dust maps, 
reduced as described in Section~\ref{temperature};
[3] Na~\textsc{i}, \citet{melendez10};
[4] Na~\textsc{i}, this study.
}
\end{deluxetable}

We adopt the \teff\ values calculated according to the
metallicity-dependent color-\teff\ calibrations given by
\citet{casagrande10}.
We estimate \teff\ from each color by drawing $10^{4}$ 
samples from each input parameter
(magnitudes, reddening, and metallicity),
assuming Gaussian uncertainties.
We adopt a minimum
uncertainty of 0.02~mag in magnitude,
0.02~mag in $E(B-V)$, and 0.3~dex in metallicity.
Not all \teff\ values for a given star 
are independent, because some magnitudes are used 
to construct multiple colors.
The same set of input draws is employed
within each of the $10^{4}$ trials, however,
so each calculation is self-consistent.
Table~\ref{tefftab} lists the median \teff\ values predicted
by each color for each star.
The uncertainty in \teff\ from a given color
is calculated as the quadrature sum of the
standard deviation of the $10^{4}$ values of \teff\
and the uncertainty in the calibration itself.
The final \teff\ value is calculated as the
average of the individual color predictions,
weighted by their inverse-square uncertainties.
These values and their statistical uncertainties (stat.)\
are listed in the final column of
Table~\ref{tefftab}. 

\begin{deluxetable*}{lccccccc}
\tablecaption{Temperatures Predicted by Casagrande et al.\ Color Calibrations
\label{tefftab}}
\tablewidth{0pt}
\tabletypesize{\scriptsize}
\tablehead{
\colhead{Star} &
\colhead{\teff} &
\colhead{\teff} &
\colhead{\teff} &
\colhead{\teff} &
\colhead{\teff} &
\colhead{\teff} &
\colhead{\teff} \\
\colhead{} &
\colhead{($B-V$)} &
\colhead{($V-J$)} &
\colhead{($V-H$)} &
\colhead{($V-K$)} &
\colhead{($J-K$)} &
\colhead{($b-y$)} &
\colhead{(adopted)} \\
\colhead{} & 
\colhead{(K)} & 
\colhead{(K)} & 
\colhead{(K)} & 
\colhead{(K)} & 
\colhead{(K)} & 
\colhead{(K)} & 
\colhead{(K)} 
}
\startdata
\bdp\     & 6589~$\pm$~175 & 6257~$\pm$~90  & 6272~$\pm$~92 & 6324~$\pm$~77 & 6570~$\pm$~278 & 6475~$\pm$~89  & 6351~$\pm$~51 \\
\bdm\     & 6447~$\pm$~168 & 6346~$\pm$~102 & 6353~$\pm$~80 & 6402~$\pm$~64 & 6611~$\pm$~262 & 6530~$\pm$~110 & 6405~$\pm$~49 \\
\cdm\     & 6522~$\pm$~167 & 6609~$\pm$~111 & 6665~$\pm$~94 & 6629~$\pm$~82 & 6642~$\pm$~257 & 6619~$\pm$~100 & 6625~$\pm$~53 \\
\gsix\    & 6516~$\pm$~185 & 6475~$\pm$~105 & 6507~$\pm$~82 & 6447~$\pm$~69 & 6481~$\pm$~259 & 6575~$\pm$~105 & 6492~$\pm$~51 \\
\hdone\   & 6093~$\pm$~147 & 5987~$\pm$~80  & 6071~$\pm$~83 & 6062~$\pm$~57 & 6281~$\pm$~217 & 6044~$\pm$~196 & 6055~$\pm$~46 \\
\hdeight\ & 6562~$\pm$~168 & 6386~$\pm$~91  & 6424~$\pm$~75 & 6382~$\pm$~62 & 6370~$\pm$~215 & 6495~$\pm$~94  & 6418~$\pm$~44 \\
\hdnine\  & 6088~$\pm$~145 & 6030~$\pm$~80  & 6047~$\pm$~67 & 6105~$\pm$~55 & 6299~$\pm$~207 & 6130~$\pm$~82  & 6087~$\pm$~40 \\
\enddata
\end{deluxetable*}

\begin{deluxetable}{lcccc}
\tablecaption{Stellar Parameters
\label{paramtab}}
\tablewidth{0pt}
\tabletypesize{\scriptsize}
\tablehead{
\colhead{Star} &
\colhead{\teff} &
\colhead{\logg} &
\colhead{\vt} &
\colhead{[M/H]} \\
\colhead{} &
\colhead{(K)} &
\colhead{} &
\colhead{(\kmsec)} &
\colhead{} 
}
\startdata
\bdp\    & 6351 $\pm$ 73 & 3.97 $\pm$ 0.13 & 1.70 $\pm$ 0.2 & $-$2.90 $\pm$ 0.1 \\
\bdm\    & 6405 $\pm$ 74 & 4.04 $\pm$ 0.15 & 1.60 $\pm$ 0.2 & $-$2.85 $\pm$ 0.1 \\
\cdm\    & 6625 $\pm$ 137& 4.29 $\pm$ 0.14 & 1.60 $\pm$ 0.2 & $-$3.00 $\pm$ 0.1 \\
\gsix\   & 6492 $\pm$ 103& 4.18 $\pm$ 0.21 & 1.55 $\pm$ 0.2 & $-$3.30 $\pm$ 0.1 \\
\hdone\  & 6055 $\pm$ 78 & 4.49 $\pm$ 0.13 & 1.20 $\pm$ 0.2 & $-$2.20 $\pm$ 0.1 \\
\hdeight\tablenotemark{a}
         & 6418 $\pm$ 117& 4.16 $\pm$ 0.14 & 1.50 $\pm$ 0.2 & $-$2.25 $\pm$ 0.1 \\
\hdnine\ & 6087 $\pm$ 84 & 4.37 $\pm$ 0.13 & 1.10 $\pm$ 0.2 & $-$1.60 $\pm$ 0.1 \\
\enddata
\tablenotetext{a}{Rederived; see Section~\ref{comparemodel}}
\end{deluxetable}

\begin{deluxetable}{lcc}
\tablecaption{Derived Fe Abundances
\label{irontab}}
\tablewidth{0pt}
\tabletypesize{\scriptsize}
\tablehead{
\colhead{Star} &
\colhead{[Fe~\textsc{i}/H]~$\pm\sigma_{\rm stat.}\pm\sigma_{\rm sys.}$ (all lines)} &
\colhead{N$_{\rm Fe~I}$} \\
\colhead{} &
\colhead{[Fe~\textsc{i}/H]~$\pm\sigma_{\rm stat.}\pm\sigma_{\rm sys.}$ (E.P.\,$>$\,1.2~eV)} &
\colhead{N$_{\rm Fe~I}$} \\
\colhead{} &
\colhead{[Fe~\textsc{ii}/H]~$\pm\sigma_{\rm stat.}\pm\sigma_{\rm sys.}$} &
\colhead{N$_{\rm Fe~II}$} \\
\colhead{} &
\colhead{[Fe~\textsc{ii}/H]$-$[Fe~\textsc{i}/H]~$\pm\sigma_{\rm stat.}\pm\sigma_{\rm sys.}$} &
\colhead{} 
}
\startdata
\bdp\    & $-$2.89 $\pm$ 0.01 $\pm$ 0.07 & 230 \\
         & $-$2.91 $\pm$ 0.01 $\pm$ 0.06 & 100 \\
         & $-$2.93 $\pm$ 0.02 $\pm$ 0.07 &  58 \\
         & $-$0.02 $\pm$ 0.02 $\pm$ 0.07 &\nodata\\
\hline
\bdm\    & $-$2.84 $\pm$ 0.01 $\pm$ 0.09 & 243 \\
         & $-$2.85 $\pm$ 0.01 $\pm$ 0.06 & 100 \\
         & $-$2.85 $\pm$ 0.02 $\pm$ 0.08 &  63 \\
         & $+$0.00 $\pm$ 0.02 $\pm$ 0.07 &\nodata\\
\hline
\cdm\    & $-$2.89 $\pm$ 0.01 $\pm$ 0.14 & 194 \\
         & $-$2.98 $\pm$ 0.01 $\pm$ 0.10 &  61 \\
         & $-$3.07 $\pm$ 0.02 $\pm$ 0.07 &  55 \\
         & $-$0.10 $\pm$ 0.02 $\pm$ 0.08 &\nodata\\
\hline
\gsix\   & $-$3.26 $\pm$ 0.02 $\pm$ 0.11 &  98 \\
         & $-$3.28 $\pm$ 0.02 $\pm$ 0.08 &  26 \\
         & $-$3.42 $\pm$ 0.02 $\pm$ 0.08 &  37 \\
         & $-$0.12 $\pm$ 0.02 $\pm$ 0.09 &\nodata\\
\hline
\hdone\  & $-$2.14 $\pm$ 0.01 $\pm$ 0.08 & 261 \\
         & $-$2.17 $\pm$ 0.01 $\pm$ 0.06 & 166 \\
         & $-$2.20 $\pm$ 0.03 $\pm$ 0.07 &  36 \\
         & $-$0.03 $\pm$ 0.03 $\pm$ 0.06 &\nodata\\
\hline
\hdeight\tablenotemark{a}
         & $-$2.24 $\pm$ 0.01 $\pm$ 0.10 & 260 \\
         & $-$2.26 $\pm$ 0.01 $\pm$ 0.09 & 164 \\
         & $-$2.23 $\pm$ 0.02 $\pm$ 0.07 &  27 \\
         & $+$0.03 $\pm$ 0.01 $\pm$ 0.07 &\nodata\\
\hline
\hdnine\ & $-$1.54 $\pm$ 0.01 $\pm$ 0.09 & 188 \\
         & $-$1.56 $\pm$ 0.01 $\pm$ 0.08 & 139 \\
         & $-$1.65 $\pm$ 0.03 $\pm$ 0.08 &  20 \\
         & $-$0.09 $\pm$ 0.03 $\pm$ 0.07 &\nodata\\
\enddata
\tablecomments{The differences in the fourth line listed for each star
reflect the [Fe~\textsc{i}/H] ratios computed when
lines with E.P.~$<$~1.2~eV are excluded.}
\tablenotetext{a}{Rederived; see Section~\ref{comparemodel}}
\end{deluxetable}

We estimate the systematic uncertainty (sys.)\ in \teff\
using the standard deviation of the average \teff\ values
predicted by this color-\teff\ scale
and those of \citet{alonso99} and \citet{ramirez05}.
These other scales provide calibrations for only
four (\citeauthor{alonso99})\ and five (\citeauthor{ramirez05})
of the colors available to us.
The total uncertainties,
which represent the quadrature sum of the statistical and systematic
uncertainties, are
reported in Table~\ref{paramtab}.

\subsection{Surface Gravity}
\label{gravity}

We calculate \logg\
from fundamental relations:\
\begin{eqnarray}
\log g = 4 \log \teff + \log(M/\msun) - 10.61 + 0.4(BC_{V}
  \nonumber\\
  + m_{V} - 5\log d + 5 - 3.1 E(B-V) - M_{\rm bol,\odot}).
\end{eqnarray}
Here, $M$ is the mass of the star, which is
always assumed to be 0.8~$\pm$~0.2~\msun.
$BC_{V}$ is the bolometric correction in the $V$ band
\citep{casagrande14}.
$m_{V}$ is the apparent $V$ magnitude.
$d$ is the distance in pc, which is
calculated from parallaxes reported in 
the first \textit{Gaia} 
data release (DR1; \citealt{gaia16}).
These values make use of the combined \textit{Tycho-2} and
\textit{Gaia} catalogs.
\textit{Gaia} DR1 did not report a parallax for \hdeight\
(Section~\ref{comparemodel}),
so we use the parallax measured 
by \textit{Hipparcos}
using the data reduction by \citet{vanleeuwen07}.
This value, 13.74~$\pm$~0.78~mas,
is slightly larger than the one measured using the 
Fine Guidance Sensors on \textit{HST} by \citet{vandenberg14},
12.24~$\pm$~0.20~mas.
The \logg\ value calculated by \citeauthor{vandenberg14},
4.05, falls within the uncertainties of the one
calculated from the \textit{Hipparcos} parallax,
4.16~$\pm$~0.14.
For the other stars in our sample, 
the \textit{Gaia} and \textit{Hipparcos} parallaxes 
agree to within $\approx$~1.5 times 
their stated uncertainties,
and the \textit{Gaia} uncertainties are typically
smaller by $\approx$~70--90\%.
$M_{\rm bol,\odot}$ is the solar bolometric magnitude, 4.75.
The constant 10.61 is calculated from the solar constants
$\log \teff_{\odot} =$~3.7617 and $\log g_{\odot} =$~4.438.
We draw $10^{4}$ samples from each
of these input parameters 
to estimate the uncertainty in \logg.
The \logg\ value for each star
reported in Table~\ref{paramtab} represents the median of these 
realizations, and
the uncertainty in \logg\ reported in Table~\ref{paramtab} is their
standard deviation.

We also compare the abundances
derived from the pressure-sensitive wings of Mg~\textsc{i} and \textsc{ii}
lines as an independent check of the \logg\ values.
The abundances derived from syntheses of
the two species should agree if the
\logg\ value is approximately correct.
The Mg~\textsc{i} line at 5183.60~\AA\ has broad wings
in only two stars in our sample, \hdone\ and \hdnine.
STIS spectra cover the Mg~\textsc{i} resonance line at 2852.13~\AA\
(except in \gsix)
and the Mg~\textsc{ii} resonance doublet at 
2795.53 and 2802.71~\AA.~
The \loggf\ values for all four of these lines are known to 
better than 3\% \citep{kramida17}, and 
\citet{barklem00} 
calculated damping constants for these lines.
The abundance uncertainties are typically 0.05--0.10~dex
and are dominated by uncertainties in the continuum placement
around the broad ($\approx$~15--30~\AA) Mg~\textsc{ii} lines.
The Mg abundances 
derived from profile fits
agree in all cases, which
supports our adopted \logg\ values.

\begin{figure}
\begin{center}
\includegraphics[angle=0,width=3.4in]{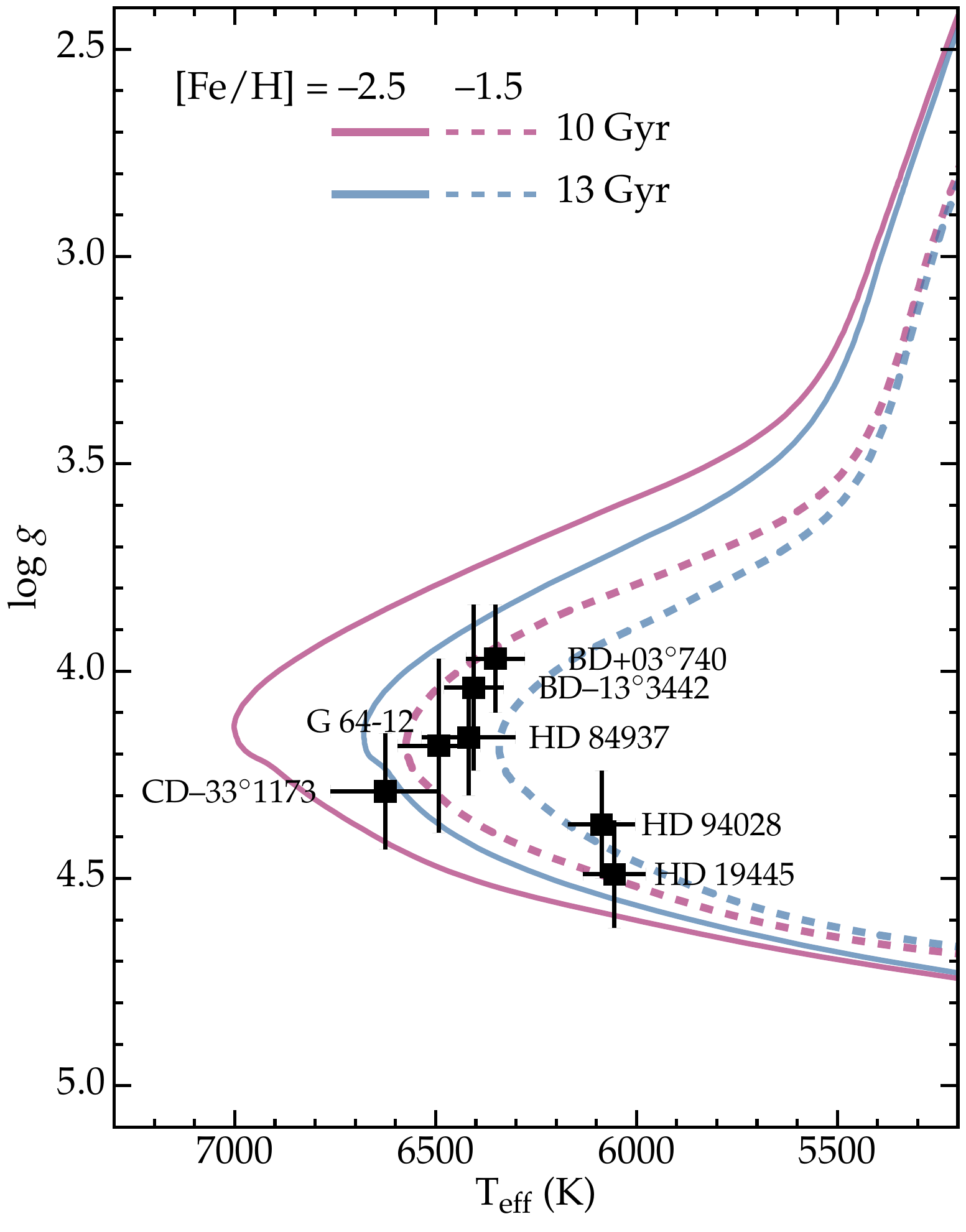}
\end{center}
\caption{
\label{isoplot}
Comparison of our calculated \teff\ and \logg\
with a set of Dartmouth isochrones \citep{dotter08}.
The colors and line types distinguish the different 
ages and metallicities, which are indicated.
The six stars in our sample,
and our calculated values for \mbox{HD~84937} (Section~\ref{comparemodel}),
are marked and labeled.
}
\end{figure}

Figure~\ref{isoplot} compares the \teff\ and \logg\
for the six stars in our sample
with a set of isochrones for 
old (10 and 13~Gyr), metal-poor 
([Fe/H]~$= -$1.5 and $-$2.5),
$\alpha$-enhanced ([$\alpha$/Fe]~$= +$0.4)
stellar populations with standard He mass fractions
(0.2452 and 0.2468)
downloaded from the Dartmouth Stellar Evolution Database \citep{dotter08}.
Several of these stars appear to lie along isochrones
that are $\sim$~1~dex more metal rich or those that would be
several Gyr older
than the age of the Universe.
Both scenarios are clearly unrealistic.
The same effect occurs if the PARSEC isochrones
\citep{bressan12,marigo17} 
are used instead.
Resolving this matter is beyond the scope of the present study,
and we underscore that the main conclusion we draw from
Figure~\ref{isoplot} is that
the stars in our sample are found on the main sequence 
or just slightly evolved beyond it.

\subsection{Microturbulent Velocity, Metallicity, and Fe Abundances}
\label{metallicity}

We interpolate model atmospheres from the $\alpha$-enhanced
ATLAS9 grid of models \citep{castelli03},
using an interpolation code provided by
A.\ McWilliam (2009, private communication).  
We derive abundances
using a recent version of the 
line analysis software MOOG
(\citealt{sneden73,sobeck11}; 2017 version).
These calculations assume that LTE holds
in the line-forming layers of the atmosphere.
This version of MOOG can calculate the contribution of
Rayleigh scattering to the continuous opacity
either as pure absorption or as isotropic, coherent scattering,
and we investigate the impact of these methods in 
Section~\ref{rayleigh}.
We use the damping constants from \citet{barklem00} 
and \citet{barklem05b}, when available. 
Otherwise, we resort to the standard \citet{unsold55} recipe.
We exclude strong lines with 
$\log({\rm EW}/\lambda) > -$4.4
from consideration.

We begin 
by adopting a microturbulent velocity of 
1.5~\kmsec\
and model metallicity consistent with
previous analyses of each star.  
We iteratively determine \vt, [M/H], and the Fe abundances
derived from Fe~\textsc{i} and \textsc{ii} lines.
During these iterations, we cull lines
whose abundance deviates by more than 0.4~dex from the
mean for each species.
We set \vt\ 
when there is no dependence of abundance derived from Fe~\textsc{i}
lines on line strength.
We set [M/H] to be the average Fe
abundance derived from Fe~\textsc{i} and \textsc{ii} lines,
rounded to the nearest 0.05~dex.
Small changes in [M/H] have minimal impact on the derived abundances.
The \vt\ and [M/H] values are
listed in Table~\ref{paramtab}, and 
the derived Fe abundances 
and number of lines used 
are listed in Table~\ref{irontab}.

We estimate systematic uncertainties on
the metallicities derived from Fe~\textsc{i} lines,
Fe~\textsc{ii} lines, and their difference 
by drawing $10^{3}$ 
samples from each input parameter in the model atmosphere
(\teff, \logg, \vt, and [M/H]) from normal distributions
using the mean values and their uncertainties
listed in Table~\ref{paramtab}.
A new model atmosphere is interpolated
for each of these $10^{3}$ draws, 
and the abundances are recomputed for each line.
We also include an EW uncertainty of 5\% for each line in these
calculations.
We adopt the median of these $10^{3}$ realizations
as the average Fe abundance
and the standard deviation 
as the systematic uncertainty.

\subsection{Comparison between Abundances Derived from 
Spectra with Different Resolving Powers}
\label{abundresolution}

We showed in Section~\ref{highresolution} that
the EWs measured from STIS E230M $R \sim$~30,000 and 
E230H $R \sim$~114,000
are statistically identical for \hdeight\ and \hdnine.
We now compare the abundances derived from these two sets
of EW values.
The two sets of EWs yield a difference
(in the sense of E230M minus E230H)
of $+$0.014~$\pm$~0.016 ($\sigma =$~0.15, 93~lines)
in \hdeight\ and
$-$0.019~$\pm$~0.060 ($\sigma =$~0.17, 8~lines) in \hdnine.
Unsurprisingly, neither difference is significant.

\subsection{Comparisons between Abundances Derived from
EWs and Spectrum Synthesis}
\label{compareewsynth}

\citet{sneden16} derived abundances in \hdeight\ by 
spectrum synthesis,
whereas we use EW-based abundances.
To compare the two approaches,
we measure EWs 
for the lines listed in Table~3 of \citeauthor{sneden16}\
from the UVES spectra of \hdeight.
We do not measure EWs for all of these lines
because some minor blends are clearly visible.
Adopting the same model atmosphere as \citeauthor{sneden16}\ used
for \hdeight,
our EW measurements
yield metallicities of
$-$2.34~$\pm$~0.01 (stat.)\ $\pm$~0.09 (sys.)\
from 162 Fe~\textsc{i} lines with E.P.~$>$~1.2~eV, and
$-$2.32~$\pm$~0.02 (stat.)\ $\pm$~0.10 (sys.)\ from 27 Fe~\textsc{ii} lines.
For the same subset of lines, the synthesis approach
yields metallicities from Fe~\textsc{i} and Fe~\textsc{ii} lines of
$-$2.30~$\pm$~0.01 (stat.)\ $\pm$~0.07 (sys.)\ and
$-$2.31~$\pm$~0.01 (stat.)\ $\pm$~0.06 (sys.), respectively.
These values are identical to the abundances derived from
the full set of Fe~\textsc{i} and Fe~\textsc{ii} lines
examined by \citeauthor{sneden16}:\
$-$2.30~$\pm$~0.01 ($\sigma =$~0.07, 446~lines) and
$-$2.31~$\pm$~0.01 ($\sigma =$~0.06, 105~lines), respectively.

We discussed in Sections~\ref{highresolution} and \ref{abundresolution}
how the
use of STIS E230M ($R \sim$~30,000) spectra compared with
higher resolution STIS E230H ($R \sim$~114,000) spectra.
We now extend that comparison to 
synthesis-based abundances derived from the E230M and E230H spectra.
We create line lists by substituting
laboratory \loggf\ values and hyperfine splitting structure
line component patterns into the
\citet{kurucz11} line lists,
and we synthesize these using MOOG.~
Figure~\ref{specplot} compares 
these synthetic spectra to the observed ones.
Many of the lines are identified, and the strengths
of most are fit reasonably well.
Improved empirical and theoretical calculations
(see, e.g., \citealt{peterson15,peterson17})
and laboratory measurements
(see, e.g., \citealt{lawler17})
are major contributors to this success.
A fair number of observed features remain
unidentified in our line lists
(e.g., lines at 
2492.24, 2496.06, and 2505.43~\AA),
and the oscillator strengths of other identified lines
are still far from precise
(e.g., lines at 
2480.39, 2492.87, and 2498.70~\AA).~
No \loggf\ values determined from
modern laboratory work have been published for these lines to the 
best of our knowledge.
We underscore that our goal in the present study is to 
analyze the abundances derived from a limited number of well-selected
lines, not provide an exhaustive analysis of all UV features.
Table~\ref{ewhighrestab} lists the lines with
abundances derived by spectrum synthesis.
The E230M spectrum and E230H spectrum of \hdeight\
yield synthesis-based abundances different by 
$-$0.017~$\pm$~0.010~dex ($\sigma =$~0.10~dex, 93~lines),
and the difference is
$-$0.055~$\pm$~0.037~dex ($\sigma =$~0.10~dex, 8~lines)
for \hdnine.
Neither of these differences is significant.

Finally,
we compare the 
EW-based and synthesis-based 
abundances derived from the E230H spectra of \hdeight\ and \hdnine.
These abundances are listed in Table~\ref{ewhighrestab}.
The difference
(in the sense of EW-based abundance minus synthesis-based abundance)
is $+$0.033~$\pm$~0.014~dex ($\sigma =$~0.14~dex, 93~lines)
in \hdeight.
The difference is
$-$0.041~$\pm$~0.039~dex ($\sigma =$~0.11~dex, 8~lines)
in \hdnine.
Again, neither of these is significant.

Several points are worth mentioning.
First, 
the differences between abundances derived from EW 
or synthesis are small ($<$~0.04~dex),
regardless of whether optical or UV spectra are considered.
Second, the standard deviation increases slightly
(by 0.02--0.04~dex)
when using EWs when compared with synthesis,
but the statistical errors on the mean remain small 
(0.01--0.02~dex)
with both techniques.
Third, the offset between the abundances derived from
Fe~\textsc{i} and Fe~\textsc{ii} lines is always 
0.02~dex or less,
regardless of whether EW-based or synthesis-based 
abundances are used.
Fourth, \hdeight\ and \hdnine\ are two of the 
three most metal-rich stars in our sample,
so any effects of blends are likely to be 
diminished in the more metal-poor stars.
We conclude from these tests
that it is acceptable to use EWs to derive Fe abundances
from these Fe~\textsc{i} and Fe~\textsc{ii}
lines, and the results are not
substantially different from those derived 
by the spectrum synthesis method.

\subsection{Other Potential Sources of Error}
\label{othererror}

Other factors in the calculations 
could potentially affect the derived
abundances. 
These include 
our choice of model atmosphere grid,
Fe partition functions, 
Fe~\textsc{ii} \loggf\ values,
and 
treatment of Rayleigh scattering in the blue and near-UV
portions of the spectrum.
We now quantify the impact of
these decisions, and we conclude that
none of these effects significantly affects
our derived [Fe/H] ratios.
We also investigate whether undetected binary companions
could bias the photometry of these stars,
thus impacting the stellar parameters we calculate.

\subsubsection{Choice of Model Atmosphere Grid}

The MARCS grid of 1D, LTE, plane-parallel model atmospheres
\citep{gustafsson08}
offers an alternative to the ATLAS9 grid.
We interpolate a set of models from this grid
with the same stellar parameters as given in Table~\ref{paramtab},
using an interpolation code provided by A.\ McWilliam
(2009, private communication).
We verify that the [Fe/H] ratios derived from either
grid of models agree to better than 0.01~dex,
so the choice of model atmosphere grid has
no impact on our derived abundances.

\subsubsection{Fe Partition Functions}

In principle,
incomplete partition functions could introduce 
an offset between the [Fe/H] ratios
derived from Fe~\textsc{i} and 
Fe~\textsc{ii} lines.
The standard 2017 version of MOOG uses 
interpolations of Fe~\textsc{i}, \textsc{ii}, and
\textsc{iii} partition functions from ATLAS9.
The derived Fe abundances change by only $-$0.002~dex
if we instead adopt interpolations of these partition
functions from the latest version (v.\ 5.5.1) of the NIST ASD.
The only unobserved levels for Fe~\textsc{i} or \textsc{ii} 
are all high, and their effects on the partition functions
will be $\lesssim$~1\%.
The ATLAS9 partition functions 
for the species of interest
are more than sufficient to compute 
Fe abundances to better than 0.01~dex precision.

\subsubsection{Fe~\textsc{ii} \loggf\ values}
\label{fe2loggf}

\citet{melendez09} presented an alternative set of 
\loggf\ values for Fe~\textsc{ii} lines in the optical part of the spectrum
($\lambda >$~4087~\AA).~
There are only a few lines that overlap with those measured by us
(3--13~lines per star).
The \citeauthor{melendez09} \loggf\ values are consistently smaller
than the NIST \loggf\ values for these lines, leading to
increases of 0.09 to 0.11~dex
in the mean abundances derived from Fe~\textsc{ii} lines.
We regard this as a potential systematic uncertainty in the
[Fe~\textsc{ii}/H] zeropoint.
Most of the Fe~\textsc{ii} lines examined by us are at shorter wavelengths
than those included in the \citeauthor{melendez09} study,
so we retain the NIST set of \loggf\ values.

\subsubsection{Treatment of Rayleigh Scattering}
\label{rayleigh}

The continuous opacity is dominated by bound-free 
transitions of the H$^{-}$ ion 
in the blue and UV portions of the spectrum
in warm, metal-poor stellar atmospheres.
The contributions from
Rayleigh scattering become significant
in cooler atmospheres,
and incorrect treatment of this contribution
can lead to overestimates of the abundances derived
from lines in the blue and UV.
Recent versions of MOOG have included the capability to 
calculate this contribution either as
pure absorption (the traditional approach)
or as isotropic, coherent scattering \citep{sobeck11}.
We verify that both approaches give identical results,
to better than 0.01~dex, in the mean [Fe/H] ratios
derived for all of the stars in our sample.
Thus the treatment of Rayleigh scattering in MOOG
has no impact on our derived abundances.

\subsubsection{Undetected Binary Companions}
\label{binaries}

The presence of undetected binary companions could 
bias the color-\teff\ calibrations by subtly altering
the input photometry.
We regard this possibility as unlikely because 
all six stars in our sample, and \hdeight,
have been subjected to long-term radial velocity monitoring
with precisions $\approx$~1~\kmsec\ or better.
These measurements span decades, yet no significant velocity
variations have been detected in any of these seven stars
(e.g., \citealt{smith98,ryan99,carney01,latham02,%
aoki06,asplund06,roederer14}).
While the presence of face-on orbits or extremely
long periods cannot be excluded, there is no 
evidence from velocities
of binarity among the stars in our sample.

\section{Comparison with Our Group's Previous Study of HD~84937}
\label{hd84937}

We now recalculate the stellar parameters and rederive the
metallicity of \hdeight\ using the methods described in 
Section~\ref{params}
and compare with the results of \citet{sneden16}.
This comparison will help establish the reliability of
our methods and place both studies on a single abundance scale.

\subsection{Comparison of Stellar Parameters and Fe Abundances}
\label{comparemodel}

We re-derive the stellar parameters of \hdeight\
using the procedure described in Section~\ref{params}.
We find 
\teff~$=$~6418~$\pm$~117~K, 
\logg~$=$~4.16~$\pm$~0.14,
\vt~$=$~1.50~$\pm$~0.20~\kmsec, and
[M/H]~$= -$2.25~$\pm$~0.10.
For comparison, \citet{sneden16} 
adopted the parameters found by \citet{lawler13},
\teff~$=$~6300~$\pm$~100~K,
\logg~$=$~4.0~$\pm$~0.2,
\vt~$=$~1.50~$\pm$~0.25~\kmsec, and
[M/H]~$= -$2.15~$\pm$~0.10.
Using our model parameters,
we derive
[Fe/H]~$= -$2.26~$\pm$~0.01 (stat.)\ $\pm$~0.09 (sys.)\ from
162 Fe~\textsc{i} lines with E.P.~$>$~1.2~eV and
[Fe/H]~$= -$2.23~$\pm$~0.02 (stat.)\ $\pm$~0.07 (sys.)\ from
27 Fe~\textsc{ii} lines.
There is a small metallicity offset between the two
analyses, mainly driven by the difference
in \teff.
This difference will be minimal when considering
the ratios of abundances of elements within the iron group,
whose line strengths respond in similar ways to changes in the
model atmosphere.

We place the \citet{sneden16} study of \hdeight\ on the same metallicity
scale as the present study by conducting a line-by-line differential analysis.
The \teff, \logg, and [Fe/H] of \hdeight\ are most similar to \bdm,
and a line-by-line differential analysis of 242
Fe~\textsc{i} lines in \hdeight\ and
\bdm\ reveals that \hdeight\ is 
0.53~$\pm$~0.01~dex ($\sigma =$~0.11~dex)
more metal-rich than \bdm.
Thus, on our scale where \bdm\ is the reference,
\hdeight\ has a metallicity of 
$-$2.32.

\subsection{External Comparison to the \textit{Gaia}-ESO Survey}
\label{ges}

\citet{lawler13} assessed the agreement of derived
stellar parameters for \hdeight\ among 
previous studies in the literature. 
\hdeight\ is the only star in our sample that
is also a benchmark star for the 
\textit{Gaia}-ESO Survey,
so we now compare those results with our own.
The \teff, \logg, and \vt\ values derived by us
are in good agreement with those derived
by \citet{jofre14} and \citet{heiter15}:\
\teff~$=$~6356~$\pm$~97~K,
\logg~$=$~4.06~$\pm$~0.04, and
\vt~$=$~1.39~$\pm$~0.24~\kmsec.
The LTE [Fe~\textsc{i}/H] ratio derived by
\citeauthor{jofre14}, $-$2.09~$\pm$~0.08,
is higher than our value by 0.17~$\pm$~0.12~dex.
Using the \textit{Gaia}-ESO Survey stellar parameters and our
line list and EWs reduces the discrepancy by 0.04~dex. 
The \textit{Gaia}-ESO Survey used only 20~Fe~\textsc{i} lines
with $\lambda >$~4900~\AA\ 
in \hdeight, and different \loggf\ values
among the 10~lines in common can account for another 
0.02~dex.
Using lines only at redder wavelengths would 
also increase our derived [Fe~\textsc{i}/H] ratio by 0.02~dex
(Section~\ref{balmerdip}).
Together, these effects can reconcile our derived [Fe~\textsc{i}/H] ratio
for \hdeight\ with that of the \textit{Gaia}-ESO Survey.

\section{Discussion}
\label{discussion}

\subsection{Good Agreement between Abundances Derived from
Fe~\textsc{i} and Fe~\textsc{ii} Lines}
\label{agreement}

The Fe abundances derived separately from
Fe~\textsc{i} and Fe~\textsc{ii} lines, 
listed in Table~\ref{irontab},
are in good agreement with each other
when strong lines and
Fe~\textsc{i} lines with E.P.~$<$~1.2~eV are excluded.
The \teff\ and \logg\ values 
are calculated largely independently from the spectra
(Section~\ref{params}),
so the agreement between [Fe~\textsc{i}/H] and [Fe~\textsc{ii}/H]
is not preordained by construction.
The offsets are small, with 
[Fe~\textsc{ii}/H]~$-$~[Fe~\textsc{i}/H]
ranging from $-$0.12~dex to 0.00~dex.
Using the \citet{melendez09} \loggf\ values
for a small subset of optical Fe~\textsc{ii} lines
(Section~\ref{fe2loggf})
would yield offsets in [Fe~\textsc{ii}/H]~$-$~[Fe~\textsc{i}/H]
of $-$0.02 to $+$0.09~dex.
Systematic uncertainties
that account for errors in the model atmosphere parameters
and EW measurements are 0.07--0.09~dex, so neither set of 
offsets is highly significant.

The Fe~\textsc{i} lines with E.P.~$<$~1.2~eV 
constitute a substantial fraction
of all Fe~\textsc{i} lines measured, ranging from 
26\% in the most metal-rich star to 73\% in the most metal-poor star.
The difference in [Fe~\textsc{i}/H] when derived from all lines
and only those with E.P.~$>$~1.2~eV is 0.03~dex or smaller in six
of the seven stars, and the difference is 0.09~dex in \cdm.
This small difference may explain why some previous studies
elected not to exclude low-E.P.\ Fe~\textsc{i} lines
(e.g., \citealt{boesgaard11}).
One of the studies that recommended omitting the low-E.P.\ lines,
\citet{lai08}, found that 
the differences in abundances derived from
the low-E.P.\ and high-E.P.\ Fe~\textsc{i} lines
were smallest in the stars with stellar parameters like 
those in our sample
(\teff~$>$~6000~K and [Fe/H]~$> -$3.3).
That study found that the average slope between the derived 
abundance and the E.P.\ differed by $-$0.01~dex~eV$^{-1}$
when all Fe~\textsc{i} lines were considered
and when only Fe~\textsc{i} lines with E.P.~$>$~1.2~eV were considered.
We adopt the set of [Fe/H] ratios derived from 
Fe~\textsc{i} lines with E.P.~$>$~1.2~eV because there
is a systematic difference, however small.
Had we chosen to retain the low-E.P.\ Fe~\textsc{i} lines,
our derived [Fe/H] ratios would be affected only minimally,
and both sets of results are presented in Table~\ref{irontab}.

Comparisons with the non-LTE corrections reported in the INSPECT database 
\citep{bergemann12,lind12}
affirm this conclusion.
The INSPECT web interface 
computes non-LTE corrections for specific Fe~\textsc{i} 
or Fe~\textsc{ii} lines by interpolating a pre-computed grid
for a given EW and stellar parameters.
There are only 15--30 Fe~\textsc{i} lines in common between the INSPECT
database and ones measured for the different stars in our study,
but
the predicted non-LTE corrections are consistent for these lines.
For each star, the non-LTE corrections for the two sets of Fe~\textsc{i} lines
(E.P.\ $<$ 1.2~eV, E.P.\ $>$ 1.2~eV)
are always consistent to within 0.03~dex.
These corrections range from $+$0.02~dex for the most metal-rich stars
to $+$0.13~dex for the most metal-poor one.
Subsequent work by \citet{amarsi16} also found consistent 
LTE and non-LTE offsets for these two sets of 
Fe~\textsc{i} lines in their 1D models,
as can be seen in their Figure~2.

Studies are in agreement that non-LTE corrections
for [Fe/H] derived from Fe~\textsc{ii} lines
are generally negligible, $\lesssim$~0.01~dex
\citep{mashonkina11,bergemann12,lind12,amarsi16}.
We verify this by checking individual Fe~\textsc{ii} lines 
in the stars in our sample with the non-LTE corrections 
reported in the INSPECT database,
and the corrections are always $<$~0.01~dex.

Our LTE [Fe/H] values are in reasonable agreement 
with other recent comparisons of LTE and non-LTE abundances
in these warm, metal-poor dwarf stars.
These studies have focused on lines available in the
optical portion of the spectrum.
\citet{mashonkina11} built a more
complete model atom for Fe than had been available previously,
and they derived 
[Fe~\textsc{ii}/H]~$-$~[Fe~\textsc{i}/H] $=$ $+$0.09~$\pm$~0.08 in LTE
for \hdeight.
When a wide range of non-LTE inelastic collision strengths
with neutral hydrogen were considered,
their non-LTE corrections to the [Fe/H] ratio derived from
Fe~\textsc{i} lines varied from $+$0.04 to $+$0.21~dex.
These non-LTE values are not in significant conflict with our LTE results.

\citet{bergemann12} studied \hdeight\ and \gsix,
and using MARCS models they found
[Fe~\textsc{ii}/H]~$-$~[Fe~\textsc{i}/H] $=$ $+$0.01~$\pm$~0.10 in LTE
for \hdeight\
and
[Fe~\textsc{ii}/H]~$-$~[Fe~\textsc{i}/H] $=$ $-$0.02~$\pm$~0.10 in LTE
for \gsix.
Their non-LTE corrections to [Fe/H] derived from 
Fe~\textsc{i} lines are 0.07~dex for \hdeight\ and
0.11~dex for \gsix.
Both the LTE and non-LTE values are in fair agreement with 
our LTE results for these two stars.

\citet{sitnova15} derived [Fe/H] from both LTE and non-LTE calculations
for 20 benchmark stars with reliable parallax measurements, 
including \hdeight\ and \hdnine.
For \hdeight\ they found 
[Fe~\textsc{ii}/H]~$-$~[Fe~\textsc{i}/H] $=$ $+$0.06~$\pm$~0.11 in LTE,
and for \hdnine\ they found
[Fe~\textsc{ii}/H]~$-$~[Fe~\textsc{i}/H] $=$ $+$0.06~$\pm$~0.11 in LTE.~
Their non-LTE corrections decreased these offsets by 0.06 and 0.02~dex, 
respectively.
These results are also in fair agreement
with our own.

Finally, \citet{ezzeddine17} predicted that the non-LTE corrections in
\gsix\ and \hdeight\ should be $+$0.29~$\pm$~0.10 and 
$+$0.14~$\pm$~0.07~dex, 
based on the extrapolation of a linear relationship
they derived from stars with [Fe/H]~$\lesssim -$4 in LTE.~
The correction for \gsix\ is notably larger than would be expected
based on our result, and the correction for 
\hdeight\ is marginally larger.
We notice that the
\citeauthor{ezzeddine17}\ linear relation changes
if only the five stars with Fe detections and
\teff~$>$~6000~K are considered:\
$\Delta$[Fe/H]~$= -$0.17$\times$[Fe/H]$_{\rm LTE}$~$-$~0.33.
This relation would predict corrections of $+$0.22 and $+$0.03~dex
for \gsix\ and \hdeight,
respectively, and the latter value is well within the 
range allowed by our results.
We emphasize that the largest Fe~\textsc{i} non-LTE corrections found by
\citeauthor{ezzeddine17}\ are predicted for stars
with lower metallicities than those considered in the present study,
and those stars are too faint for comparable UV observations with STIS.

\subsection{Abundances Derived from Lines in the Balmer Dip Region}
\label{balmerdip}

\begin{figure*}
\begin{center}
\includegraphics[angle=0,width=3.35in]{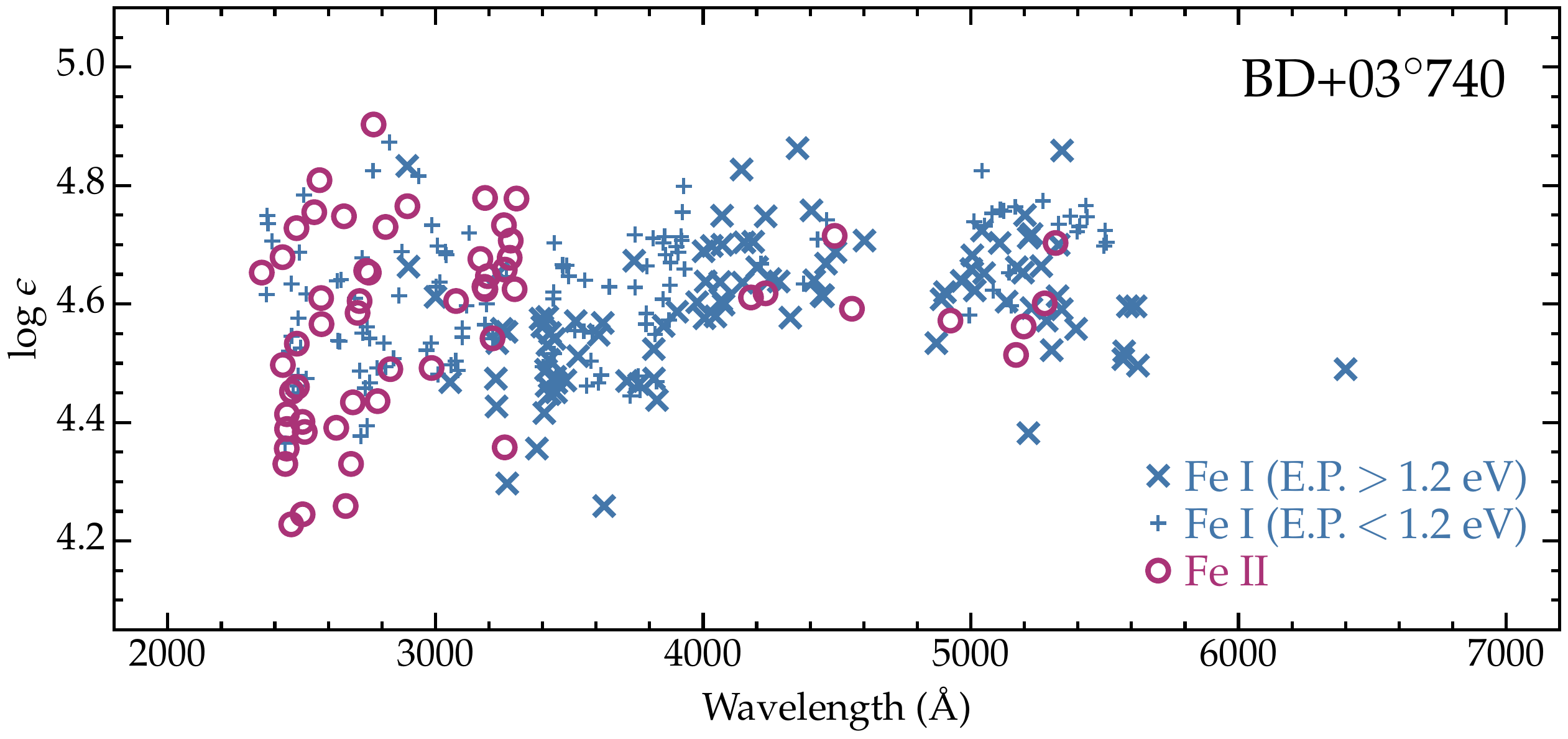}
\hspace*{0.1in}
\includegraphics[angle=0,width=3.35in]{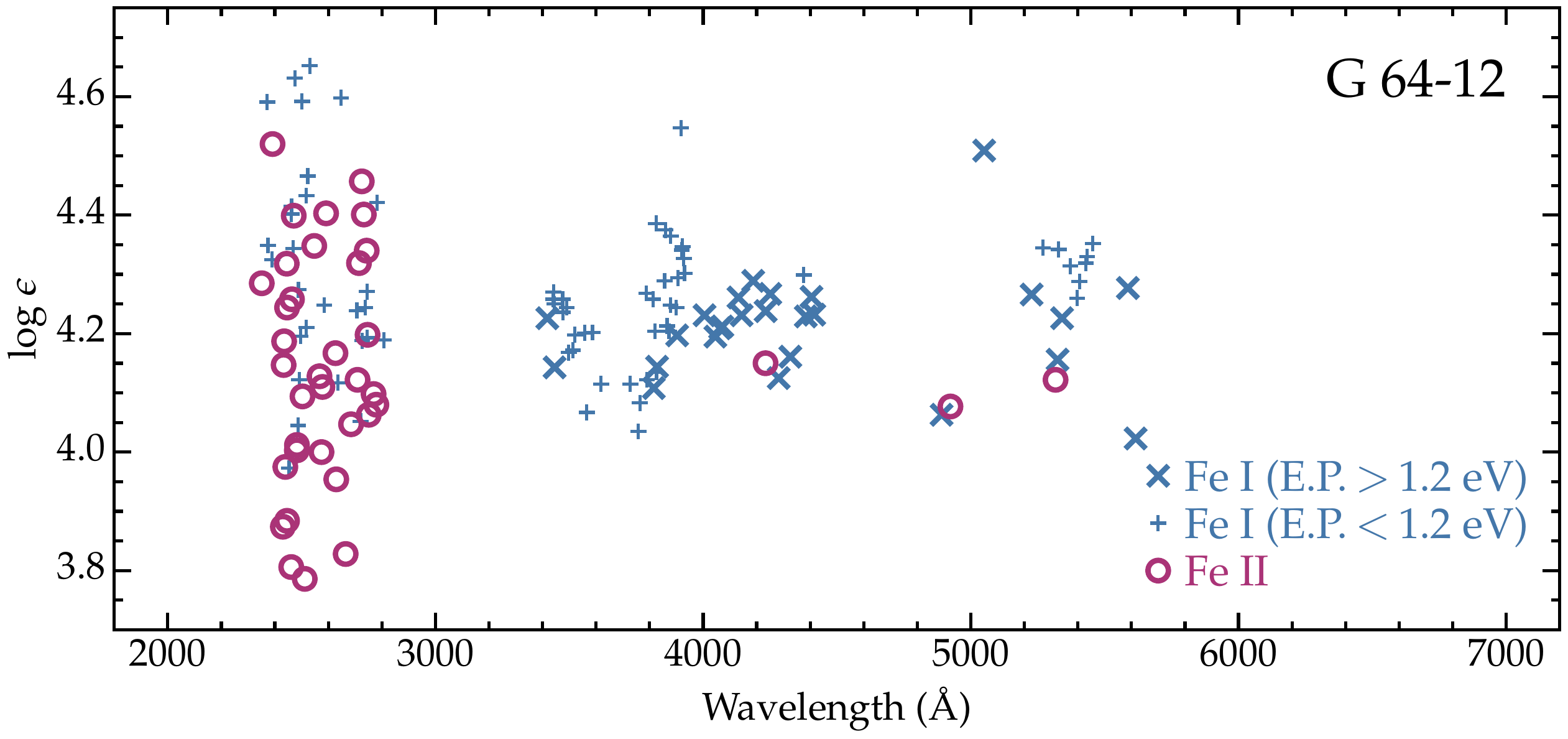} \\
\includegraphics[angle=0,width=3.35in]{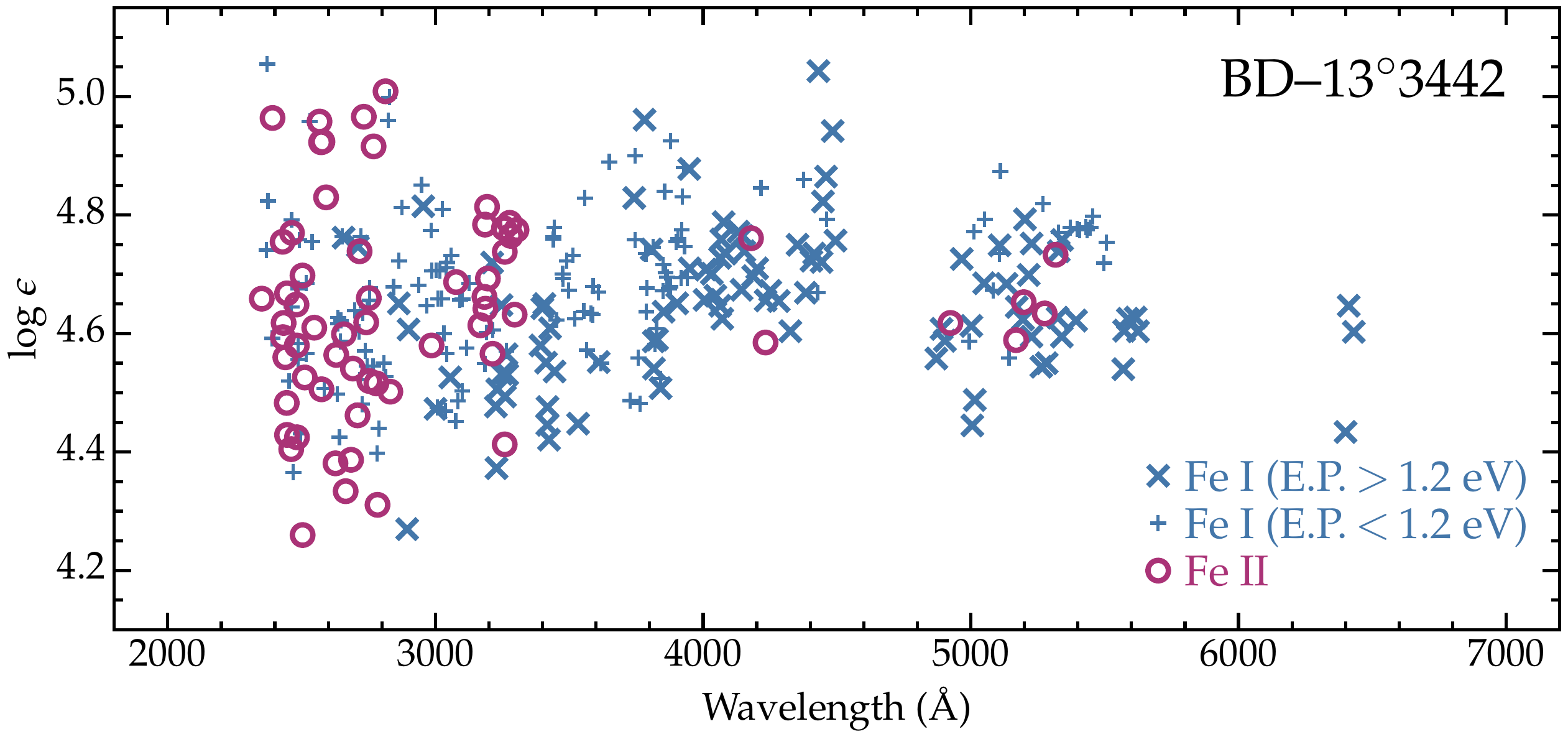}
\hspace*{0.1in}
\includegraphics[angle=0,width=3.35in]{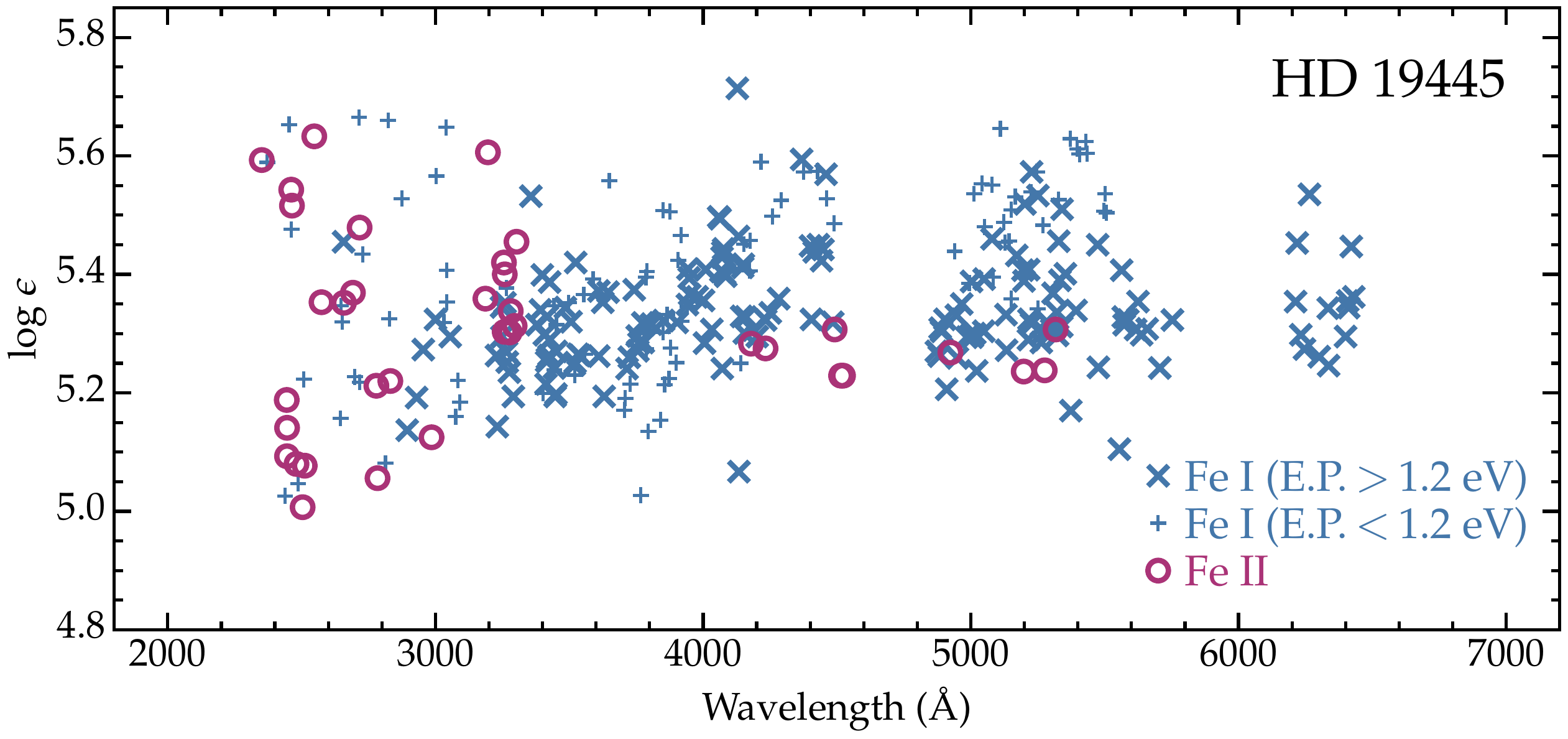} \\
\includegraphics[angle=0,width=3.35in]{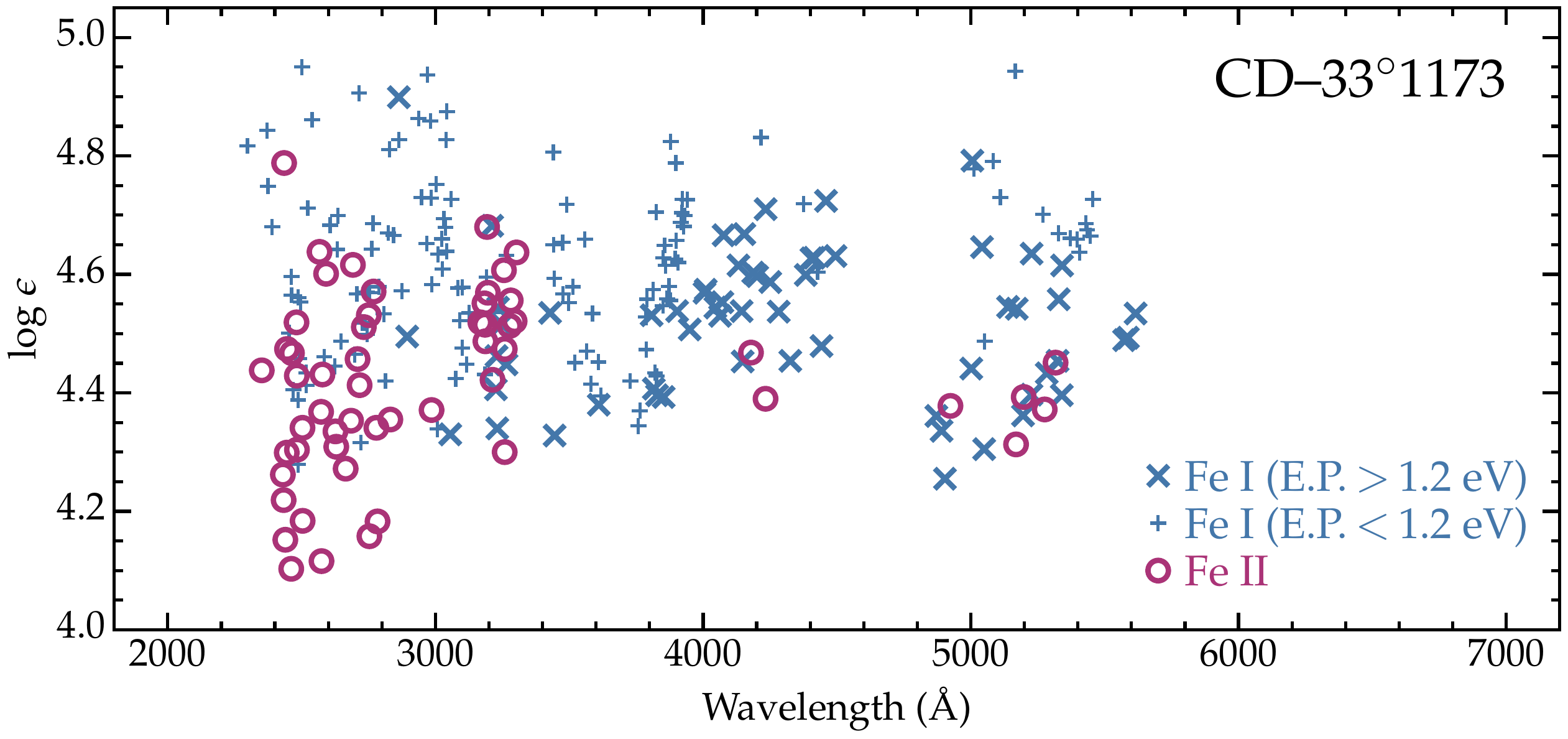}
\hspace*{0.1in}
\includegraphics[angle=0,width=3.35in]{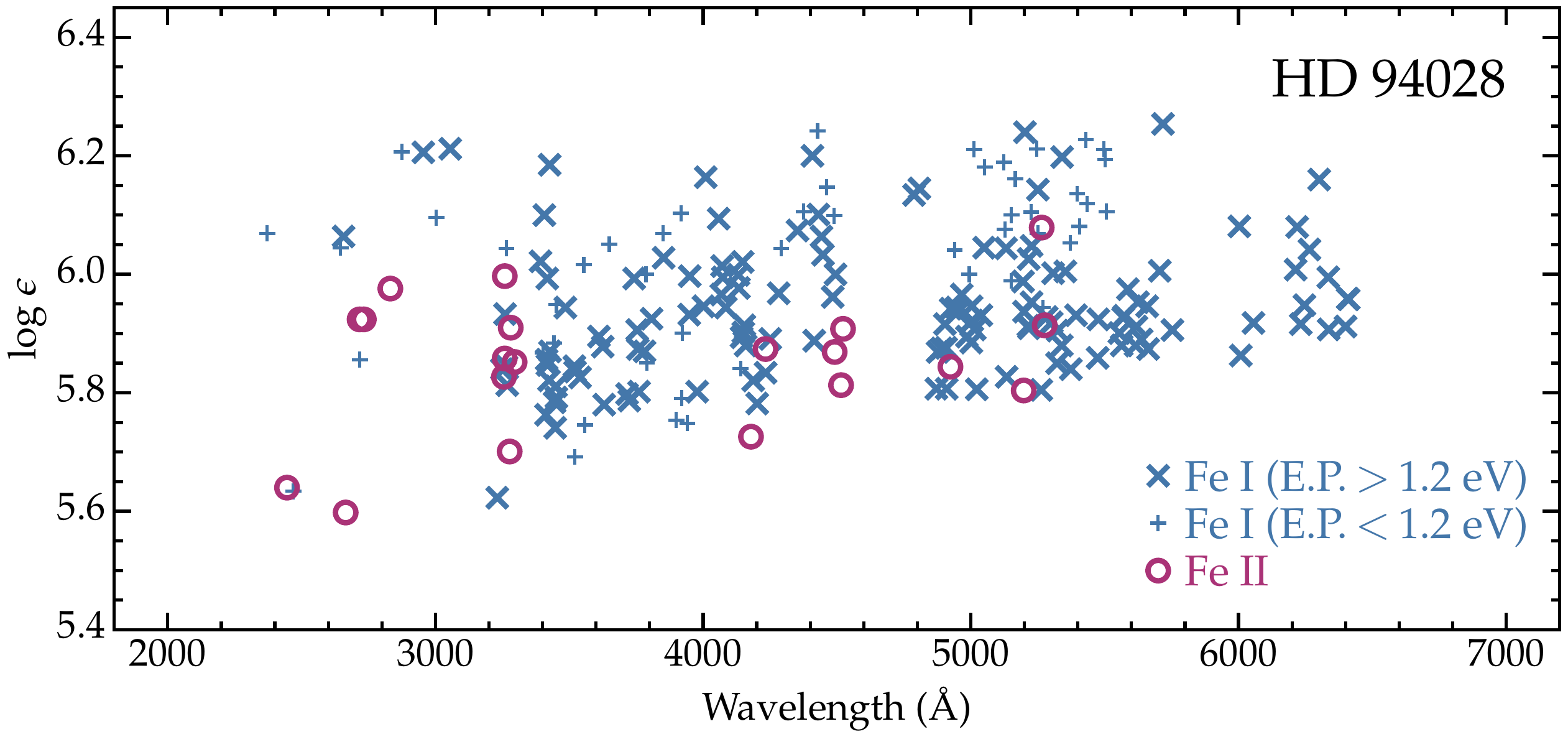}
\hspace*{3.45in}
\end{center}
\caption{
\label{plotwave}
Derived Fe abundances versus wavelength for the six
stars in our sample.
The blue crosses and plus signs denote Fe~\textsc{i} lines,
and the red circles denote Fe~\textsc{ii} lines.
}
\end{figure*}

Figure~\ref{plotwave} shows abundance trends 
as a function of wavelength.
The Fe~\textsc{i} lines between 3100 and 3700~\AA\
have mildly (0.03--0.14~dex) lower abundances, on average,
than lines at shorter or longer wavelengths.
Table~\ref{balmerdiptab} lists these values.
\citet{roederer12} noticed that Fe~\textsc{i} lines in this
wavelength range yielded abundances 
that were systematically lower by 0.16 to 0.27~dex
than those at shorter or longer wavelengths in four metal-poor giants.
Our study confirms a similar effect in the six dwarfs studied
here, although the magnitude of the effect is substantially reduced.
The data of \citet{sneden16} also reveal 
a reduction of 0.08~dex in the abundances 
derived from Fe~\textsc{i} lines with 3100--3700~\AA\ in \hdeight,
in agreement with the sign and magnitude of our results
for other dwarf stars.
Hereafter, we refer to this as the ``Balmer Dip'' effect,
reflecting the fact that it affects lines
near the convergence of the Balmer series and the
beginning of the Balmer continuum wavelength region.

\begin{deluxetable*}{lcccccc}
\tablecaption{Fe Abundance Deviations in the 3100--3700~\AA\ Region
\label{balmerdiptab}}
\tablewidth{0pt}
\tabletypesize{\scriptsize}
\tablehead{
\colhead{} &
\multicolumn{2}{c}{$\lambda <$~3100 \AA, $\lambda >$~3700 \AA} &
\colhead{} &
\multicolumn{2}{c}{3100 $\leq \lambda \leq$ 3700 \AA} &
\colhead{} \\
\cline{2-3} \cline{5-6}
\colhead{Star} &
\colhead{$\langle \log\epsilon \rangle$} &
\colhead{N} &
\colhead{} &
\colhead{$\langle \log\epsilon \rangle$} &
\colhead{N} &
\colhead{$\Delta$} 
}
\startdata
\multicolumn{7}{c}{Abundances from Fe~\textsc{i} lines} \\
\hline
\bdp\    & 4.63 $\pm$ 0.01 &  72 & & 4.49 $\pm$ 0.02 & 28 & $+$0.14 $\pm$ 0.03 \\
\bdm\    & 4.67 $\pm$ 0.02 &  80 & & 4.54 $\pm$ 0.03 & 20 & $+$0.13 $\pm$ 0.03 \\
\cdm\    & 4.53 $\pm$ 0.02 &  51 & & 4.46 $\pm$ 0.04 & 10 & $+$0.07 $\pm$ 0.05 \\
\gsix\   & 4.21 $\pm$ 0.03 &  24 & & 4.18 $\pm$ 0.07 &  2 & $+$0.03 $\pm$ 0.07 \\
\hdone\  & 5.35 $\pm$ 0.01 & 129 & & 5.29 $\pm$ 0.02 & 37 & $+$0.06 $\pm$ 0.02 \\
\hdeight\tablenotemark{a}
         & 5.22 $\pm$ 0.01 & 367 & & 5.14 $\pm$ 0.01 & 79 & $+$0.08 $\pm$ 0.01 \\
\hdnine\ & 5.96 $\pm$ 0.01 & 114 & & 5.87 $\pm$ 0.03 & 25 & $+$0.09 $\pm$ 0.03 \\
\hline\hline
\multicolumn{7}{c}{Abundances from Fe~\textsc{ii} lines} \\
\hline
\bdp\    & 4.55 $\pm$ 0.03 &  45 & & 4.65 $\pm$ 0.04 & 13 & $-$0.10 $\pm$ 0.04 \\
\bdm\    & 4.63 $\pm$ 0.03 &  49 & & 4.69 $\pm$ 0.04 & 14 & $-$0.06 $\pm$ 0.05 \\
\cdm\    & 4.38 $\pm$ 0.03 &  41 & & 4.52 $\pm$ 0.03 & 14 & $-$0.14 $\pm$ 0.04 \\
\gsix\   & 4.14 $\pm$ 0.03 &  37 & & \nodata         &  0 & \nodata            \\
\hdone\  & 5.27 $\pm$ 0.03 &  27 & & 5.39 $\pm$ 0.04 &  9 & $-$0.12 $\pm$ 0.05 \\
\hdeight\tablenotemark{a}
         & 5.19 $\pm$ 0.01 &  91 & & 5.22 $\pm$ 0.01 & 14 & $-$0.03 $\pm$ 0.01 \\
\hdnine\ & 5.85 $\pm$ 0.04 &  14 & & 5.86 $\pm$ 0.05 &  6 & $-$0.01 $\pm$ 0.06 \\
\enddata
\tablenotetext{a}{Using abundances presented in \citet{sneden16}}
\end{deluxetable*}

Curiously, the effect is reversed for some of the stars when Fe~\textsc{ii}
lines are considered.
Table~\ref{balmerdiptab} also lists these values.
The number of Fe~\textsc{ii} lines 
in the 3100--3700~\AA\ region is limited, however, to a handful of 
lines with wavelengths between 3167 and 3303~\AA,
so these lines are less informative than the Fe~\textsc{i} lines, which
span the full wavelength range.
The analysis of Fe~\textsc{ii}
within the discrepant wavelength range of
\hdeight\ is also restricted to this set of Fe~\textsc{ii} lines.

The Balmer Dip effect is not unique to the MOOG line analysis code.
Similar results are clearly seen, for example, in Figure~6 of
\citet{ito13}, who found a systematic decrease 
of $\approx$~0.10--0.15~dex in abundances derived from
Fe~\textsc{i} lines in the Balmer Dip region
in the metal-poor giant (\teff~$=$~5430~K)
\object[BD+44 493]{BD$+$44$^{\circ}$493}.
That study made use of a line analysis code derived from 
the work of \citet{tsuji78}.
The analysis of the warm (\teff~$=$~6050~K) dwarf star 
\object[UCAC3 133-60515]{WISE~J0725$-$2351} by \citet{spite15}
using the Turbospectrum code \citep{alvarez98} 
reveals an abundance decrease of 0.18~$\pm$~0.03~dex
when derived from Fe~\textsc{i} lines inside the
Balmer Dip region.
We have not performed an exhaustive search for
other studies that show similar effects,
but these two examples demonstrate that the
Balmer Dip effect is found in Fe abundances derived from
multiple line analysis codes.

One possible cause of the Balmer Dip effect 
could be that lines in this wavelength range preferentially 
arise from electronic levels whose populations are 
overionized relative to their LTE values, leading 
to an underprediction of the abundances.
If so, all lines arising from these levels should
yield low abundances, whether or not the
wavelengths are in the Balmer Dip region.
\citet{wood13}, who focused on Ti~\textsc{ii}, performed
related tests for lines in \hdeight,
and concluded that this was not the cause of the effect.
In that study,
Ti~\textsc{ii} lines arising from 
the more highly-excited levels ($>$~0.6~eV)
with wavelengths in the Balmer Dip region
yielded low abundances,
but lines at other wavelengths arising from these levels
and lines inside and outside the Balmer Dip region 
arising from lower levels yielded consistent abundances.

Here, equipped with lines in six additional stars,
we are in a better position to definitively test
this hypothesis.
For a given star, we identify levels that have at least
three Fe~\textsc{i} lines 
in the Balmer Dip region 
and at least three Fe~\textsc{i} lines outside the Balmer Dip region.
For each level, we compute the mean abundance derived from
lines inside the Balmer Dip region and those outside the Balmer Dip region.
We repeat this test for each of the six stars in our sample
and \hdeight\ from \citet{sneden16}.
A difference of zero would indicate that
the level populations themselves could be the source of the Balmer Dip effect.
A non-zero difference would indicate that
the levels are not the cause of the effect.

\begin{figure}
\begin{center}
\includegraphics[angle=0,width=3.4in]{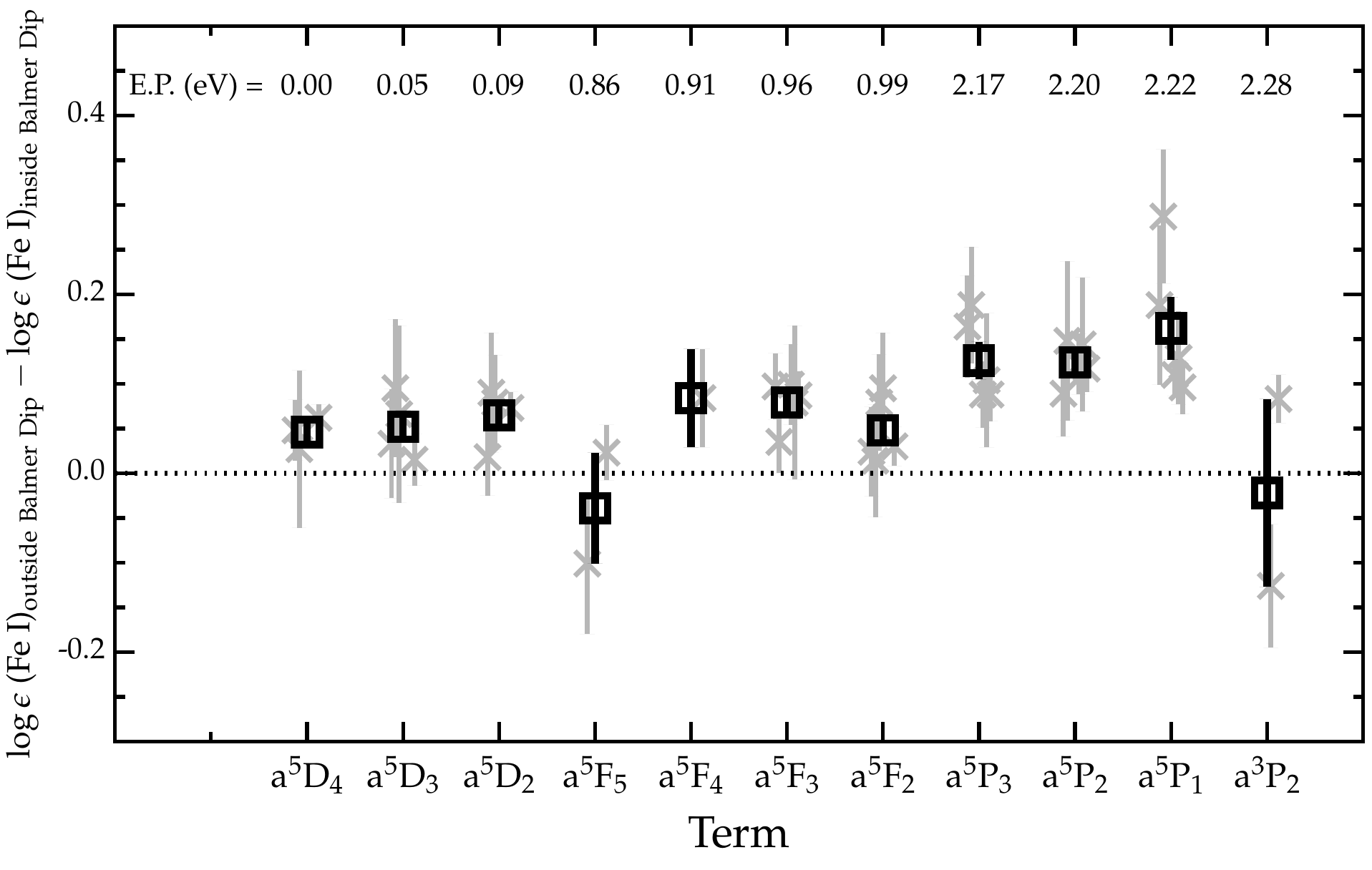}
\end{center}
\caption{
\label{balmerdipplot}
Mean abundance differences derived from Fe~\textsc{i} lines
outside and inside the Balmer Dip region 
(3100~$\leq \lambda \leq$~3700~\AA).~
The gray crosses---offset horizontally slightly for clarity---indicate 
the measurements from individual stars,
where the number of lines per level ranges from 9 to 37,
with a median of 15 lines.
The gray error bars indicate the standard error on the difference.
The black squares indicate the mean differences for a given level,
and the black error bars indicate the standard error on this difference.
The term designations 
are listed along the bottom, and the excitation potentials
are listed along the top.
The dotted line indicates zero difference.
}
\end{figure}

The results of this test are shown in Figure~\ref{balmerdipplot}.
There are 11 electronic levels of neutral Fe that 
meet our criteria, although not all levels meet these criteria
for each of the seven stars considered.
These levels are associated with multiplets from the 
$3d^{6}4s^{2}$ $a^{5}D$ ground state
and the 
$3d^{7}(^{4}F)4s$ $a^{5}F$,
$3d^{6}4s^{2}$ $a^{3}P$,
and
$3d^{7}(^{4}P)4s$ $a^{5}P$ excited states.
The mean abundance differences between
Fe~\textsc{i} lines outside the Balmer Dip region and
Fe~\textsc{i} lines inside the Balmer Dip region 
are usually positive
and non-zero, and most are significant by
several standard deviations.
Two levels are each discrepant in one star
($a^{5}F_{5}$ in \bdm\
and 
$a^{3}P_{2}$ in \hdone),
but these levels only meet our criteria for inclusion
in two stars, so we are reluctant to draw firm 
conclusions from them.
The $a^{5}F_{4}$ level only meets our criteria for inclusion
in one star (\hdeight), so we also do not draw firm
conclusions from it.
Individual levels often give rise to measured lines
at wavelengths shorter and longer than the Balmer Dip region,
so the upper levels are not influencing the results found here.
The significant, non-zero differences for the remaining
eight levels
falsify our hypothesis that the LTE level populations in 
these particular electronic levels 
could be the source of the Balmer Dip effect.

Unfortunately, no levels of singly-ionized Fe meet our criteria
in any star in the sample, 
so we cannot perform this test for Fe~\textsc{ii} lines.

The possibility that the Balmer Dip effect
could be a consequence of 
systematic uncertainties in the 
branching fractions of upper transition levels
has also been considered by \citet{lawler18}.
That study discussed several effects
that could impact the calibration of
branching ratios of Ar~\textsc{i} and \textsc{ii}
in laboratory hollow cathode lamp spectra,
which are commonly used to calibrate branching fractions of
other species.
\citeauthor{lawler18}\ conclude that
these calibration issues are not likely to be important
in the Balmer Dip effect, but 
further tests are ongoing.

Previous studies of lines of other species
have yielded a range outcomes with regard
to systematic decreases in abundances derived from
lines in the Balmer Dip region.
Lines of Ti~\textsc{i}, Ti~\textsc{ii}, and possibly
V~\textsc{ii} show the effect in \hdeight\
\citep{lawler13,lawler14,wood13,wood14b}.
Lines of Sc~\textsc{ii}, Cr~\textsc{ii},
Co~\textsc{i}, and Ni~\textsc{i} do not
show any discrepancy in \hdeight\ \citep{wood14a,lawler15,sneden16}.
Lines of V~\textsc{i}, Cr~\textsc{i}, 
Mn~\textsc{i}, and Mn~\textsc{ii} 
yield inconclusive results, because the
number of lines available in the affected wavelength
region of \hdeight\ is small \citep{lawler14,sneden16}.
These results suggest that, at least in analyses of \hdeight,
missing or incorrect treatment of the
continuous opacity in the Balmer Dip region
cannot be a major factor in the Balmer Dip effect.
\citet{lawler13} and \citet{wood13} suggested that
non-LTE effects in the H~\textsc{i} $n =$~2 level
may be (partially) responsible.
If so, 
then a closer comparison of the sets of species that are affected
and the sets of species that are not affected
may be key to diagnosing the line-forming layers of the
atmosphere where the effect originates.
As such, the Balmer Dip effect may be a useful
and readily accessible tool 
to better understand 3D convection effects
in the atmospheres of distant stars.
Further investigation of this matter with the full set of
iron-group abundances in our sample of six stars
may offer new insight here.


\section{Conclusions}
\label{conclusions}

We have collected new and archival high-resolution UV and optical 
spectra of six warm, metal-poor dwarf stars.
Using stellar parameters calculated largely independent of the
spectra themselves, we have derived [Fe/H] ratios
from several hundred Fe~\textsc{i} and Fe~\textsc{ii} lines
with wavelengths between 2290 and 6430~\AA.~
The Fe~\textsc{ii} lines should be adequately modeled
using standard LTE assumptions,
whereas Fe~\textsc{i} may not.
The [Fe/H] ratios derived separately from the two species
are in agreement with each other to within
$\approx$~1.3 times their uncertainties
when strong lines and Fe~\textsc{i} lines with E.P.~$<$~1.2~eV are excluded.
These results constrain the limits of departures from
LTE to be 
minimal, at most,
within the range of stellar parameters considered,
for the higher-excitation lines.
This is in agreement with modern studies of non-LTE line formation
in stars like these.
Theoretical calculations of Fe~\textsc{i} lines in non-LTE
predict that the departures from LTE should increase 
substantially at metallicities lower than those 
examined by our study (e.g., \citealt{ezzeddine17}),
so our results may not be generalizable to
lower-metallicity dwarfs.

We have attempted to identify possible sources of 
error in our method and calculations, including 
the choice of 1D LTE plane-parallel
model atmosphere grids,
Fe partition functions, and
the treatment of Rayleigh scattering in MOOG.~
None of these affect our derived [Fe/H] ratios
in excess of 0.01~dex.
A yet-unidentified source of uncertainty 
affects some abundances derived from lines in the
wavelength region from approximately 
3100 to 3700~\AA,
as has been found previously (cf., e.g., \citealt{roederer12}).
Our investigations indicate that this Balmer Dip effect
cannot be attributed exclusively to missing continuous opacity
or non-LTE effects in several Fe~\textsc{i} electronic levels
that give rise to lines in this wavelength region.
An analysis of the impact of the Balmer Dip effect
on additional iron-group species may help to resolve the matter.

\acknowledgments

We thank the referee for suggesting several
tests that have helped to strengthen the
conclusions of this work.
IUR also thanks A.\ Ji for useful conversations about MOOG.~
Generous support for Program GO-14232 has been provided by 
a grant from STScI, which is operated by AURA,
under NASA contract NAS5-26555.
We also acknowledge partial support from
National Science Foundation (NSF) grants 
PHY~14-30152 (Physics Frontier Center/JINA-CEE) and
AST~16-16040 (to CS).~
This research has made use of NASA's
Astrophysics Data System Bibliographic Services;
the arXiv pre-print server operated by Cornell University;
the SIMBAD and VizieR
database hosted by the
Strasbourg Astronomical Data Center;
the Atomic Spectra Database hosted by 
the National Institute of Standards and Technology;
the MAST at STScI; 
the INSPECT database 
(v.\ 1.0; \url{http://www.inspect-stars.com});
and the
IRAF software packages
distributed by the National Optical Astronomy Observatories,
which are operated by AURA,
under cooperative agreement with the NSF.
This work has also made use of data from the European Space Agency (ESA)
mission {\it Gaia},
(\url{http://www.cosmos.esa.int/gaia}),
processed by
the {\it Gaia} Data Processing and Analysis Consortium (DPAC).
\url{http://www.cosmos.esa.int/web/gaia/dpac/consortium}).
Funding for the DPAC has been provided by national institutions, in particular
the institutions participating in the {\it Gaia} Multilateral Agreement.

\facility{%
HST (STIS), 
Keck I (HIRES), 
Smith (Tull),
VLT (UVES)}

\software{%
IRAF \citep{tody93},
MARCS \citep{gustafsson08},
matplotlib \citep{hunter07},
MOOG \citep[][2017 version]{sneden73,sobeck11},
numpy \citep{vanderwalt11},
R \citep{rsoftware},
scipy \citep{jones01}}

\end{document}